# Critically Evaluated Spectral Data for Singly Ionized Carbon (C II)

A. Kramida[1,4] and K. Haris[2,3]
[1] National Institute of Standards and Technology, Gaithersburg, MD 20899, USA; alexander.kramida@nist.gov
[2] Associate, National Institute of Standards and Technology, Gaithersburg, MD 20899, USA
[3] Department of Physics, Aligarh Muslim University, Aligarh 202002, India

## Abstract

All available experimental data on the spectrum of singly ionized carbon have been critically analyzed. Measurement uncertainties of all published studies have been reassessed. The scope of observational data includes laboratory emission spectra of arcs, sparks, electrodeless discharges, and hollow cathode lamps recorded with grating and Fourier transform spectrometers, laboratory photoabsorption spectra, and emission spectra of planetary nebulae. The total number of observed spectral lines included in this compilation is 597. These lines participate in 972 transitions. From this list of identified transitions, we have derived a set of 414 energy levels, which are optimized using a least-squares fitting procedure. The identifications are supported by parametric calculations with Cowan's codes. The existing tables of critically evaluated transition probabilities have been extended with our newly calculated data. The ionization energy has been derived from the newly optimized energy levels with improved precision. Data on the isotope shifts and hyperfine structure have also been compiled.

*Unified Astronomy Thesaurus concepts:* Atomic spectroscopy (2099); Atomic physics (2063); Transition probabilities (2074); Isotope shifts (2069); Spectral line identification (2073); Line intensities (2084); Line positions (2085)

*Supporting material:* machine-readable tables

## 1. Introduction

Carbon is a chemical element constituting the basis of all known life forms. It is ubiquitous in the universe. Its atoms and ions are found in free and chemically bound forms in atmospheres of planets, stars, and many other astrophysical objects (Henning & Salama 1998). Carbon plays a key role in the evolution of stars (the so-called CNO cycle) and is also abundant in the interstellar medium (ISM). In the ISM, a large part of all carbon is in the form of $C^+$ as, due to the low ionization threshold of neutral C ($\approx$11.26 eV), carbon ions can easily be produced by photoionization. Thermal collisions of $C^+$ with lighter particles, such as $e^-$, H, and D, induce the [C II] $2s^2 2p$ $^2P^\circ$ $J = 3/2 \rightarrow 1/2$ emission line at wavelength $\lambda \approx 158\,\mu$m, which is an effective coolant of the ISM and thus plays an important role in its evolution (Wolfire et al. 2003). This line is used as a probe for variations of fundamental constants (Levshakov et al. 2017, 2020). Many other spectral lines of C II have been observed in spectra of the Sun, planetary nebulae, novae, Seyfert galaxies, and the ISM (Kaler 1976; Curdt et al. 2001; Sharpee et al. 2003, 2004; Parenti et al. 2005; García-Rojas et al. 2012; Iijima & Nakanishi 2008; Gnaciński 2009; Otsuka et al. 2010; Fang & Liu 2011; Zhang et al. 2005 and references therein).

For laboratory studies and applications, reliable atomic data on carbon are of great need, as it is one of the most extensively used and commonly found impurities in industrial plasmas and laboratory light sources. For example, wavelength measurements in the vacuum ultraviolet (VUV) region often employ carbon spectral lines for calibration. Fusion energy research requires extensive sets of atomic data for many light elements including carbon (Braams & Chung 2015). For example, lines of C II are used for diagnostics of tokamak plasmas (Isler et al. 1997; Menmuir et al. 2014). It was recently demonstrated that the erosion of heavy metals (W or Mo) by tokamak plasmas can be abated by injection of methane and deuterium gases (Ding et al. 2017); carbon lines are important for diagnostics of such plasmas. Spectral features of various atomic species including C II-V have been recorded in emission of tokamak plasmas in a wide range of wavelengths from visible to VUV (von Hellermann et al. 2005; Graf et al. 2011; Oishi et al. 2014; McCarthy et al. 2016).

Singly ionized carbon belongs to the boron (B I) isoelectronic sequence with the ground configuration [He]$2s^2 2p$ comprised of two levels, $^2P^\circ_{1/2,3/2}$, $J = 1/2$ being the lowest one. Excitation of the valence electron creates configurations of the type $2s^2 n\ell$ ($n > 2$, $\ell = s, p, d, f, g, h, ...$) with doublet ($^2L_J$) levels. The $2s2p^2$ configuration containing the $^4P$, $^2D$, $^2S$, and $^2P$ multiplets is created from the ground configuration $2s^2 2p$ by excitation of the $2s$ electron. Further excitation from $2s2p^2$ leads to the $2s2p(^{1,3}P^\circ)n\ell$ ($n > 2$, $\ell = s, p, d, f$, g, h, ...) and $2p^3$ configurations. Core-excited resonance states with a vacancy in the $K$ shell can be created by photoexcitation from the ground state, $1s^2 2s^2 2p$ $^2P^\circ \rightarrow 1s2s^2 2p^2$ $^ML_J + 1s2s^2 2pn\ell$ $^ML_J$ ($n > 3$, $\ell = s$, $p$, $d$, ...), or from the lowest metastable multiplet, $1s^2 2s2p^2$ $^4P \rightarrow 1s2s2p^3$ $^ML^\circ_J + 1s2s2p^2 n\ell$ $^ML^\circ_J$ ($n > 3$, $\ell = s$, $p$, $d$, ...).

Similar to the neutral carbon spectrum, early investigations on C II were made in the 1920s. A few of them are noteworthy (Bowen & Ingram 1926; Bowen 1927; Fowler & Selwyn 1928; see Haris & Kramida 2017 for the review of the C I spectrum). In continuation of these studies, Edlén (1933a, 1933b, 1934, 1936) remeasured the C II spectrum in a wide range of wavelengths from VUV to near-infrared and classified most of the observed lines. Edlén's analysis was the basis for the energy levels given by C. E. Moore in her famous *Atomic Energy Levels* monograph (Moore 1949). Some accurately measured wavelengths of C II were reported by Boyce & Rieke (1935) in their list of provisional wavelength standards in the VUV, which revised the less accurate wavelengths given earlier by Bowen & Ingram (1926) and by Bowen (1927). A major progress in the analysis of C II was made by Glad (1953) who recorded and accurately measured the spectrum in a wide wavelength range from 198 to 880 nm in the emission of a condensed hollow cathode (HC) discharge. A few years later, Herzberg (1958) remeasured two groups of C II lines near 176 and 133.5 nm with high precision. His wavelength uncertainties were as small as 0.00007 nm, which was achieved by measuring these lines in high orders of diffraction on a grating instrument. These advances were utilized for the redetermination of the energy levels by Edlén (1963). Using these new energy level values, he extended the list of accurate Ritz wavelengths in the VUV region. Edlén's energy levels were adopted by Moore (1970, 1993) in her compilations of carbon spectra (hereafter called Moore's tables), with an interchange of the $^4D_{5/2}$ and $^2D_{5/2}$ levels of the $2s2p(^3P°)4f$ and $2s2p(^3P°)5f$ configurations, and with a suggested value of the ionization energy (IE) of C II derived from the data of Glad (1953). Since then, a few transitions were measured in laboratory experiments with extraordinary precision. For example, the fine-structure interval in the ground term $2s^22p\ ^2P°$ was precisely measured at ≈1900 GHz in both stable isotopes $^{12}$C and $^{13}$C using the laser magnetic resonance technique (Cooksy et al. 1986). Recently, the wavelengths of the $2s^22p\ ^2P°–2s2p^2\ ^2D$ transitions were measured with a Fourier transform spectrometer (FTS) in a HC discharge (Nave & Clear 2021).

---

[4] Corresponding author.

In addition to laboratory studies, a few accurate VUV line measurements were performed in different solar experiments (Curdt et al. 2001; Parenti et al. 2005). Wavelengths of the $2s^22p\ ^2P–2s2p^2\ ^4P$ intercombination transitions were measured in the spectra of several astrophysical objects (Doschek et al. 1977; Penston et al. 1983; Young et al. 2011).

Advances in powerful lasers and synchrotron radiation sources (SRS) led to progress in the photoionization spectroscopy methods, which allowed measurements of photoabsorption features near K and L edges of C II originating from the ground term $2s^22p\ ^2P°$ and the lowest metastable term $2s2p^2\ ^4P$ (Jannitti et al. 1993; Nicolosi & Villoresi 1998; Kjeldsen et al. 2001; Recanatini et al. 2001; Müller et al. 2018). Interpretation of these spectra remains a challenging problem. This is evidenced by contradictions in the classification of many observed features reported in different papers.

Only a few experimental studies of isotopic effects have been published for the C II spectrum (Burnett 1950, 1957; Cooksy et al. 1986; Haridass & Huber 1994), whereas several accurate theoretical calculations are now available (e.g., Berengut et al. 2006; Korol & Kozlov 2007; Jönsson et al. 2010; Bubin & Adamowicz 2011). Evaluation of these theoretical data allows estimation of their uncertainties, and the best theoretical results can be adopted as provisional values for various applications.

Data on C II transition parameters, such as spontaneous radiative decay rates (A-values), oscillator strengths (f-values), line strengths (S), Stark shift, and broadening parameters are abundant in the literature. Selecting the best data for each transition from hundreds of published papers can be a daunting task. Fortunately, two sets of critically evaluated data are available: (i) Stark widths and shifts (Konjević et al. 2002), and (ii) transition probabilities (TPs; Wiese et al. 1996; Wiese & Fuhr 2007). These critical compilations provide a basis for the critical evaluation of the data published more recently. Data on Stark widths and shifts are outside the scope of the present work, but the data on TP are extensively used, and one of the aims of this work is to extend them with new A-values.

One of the most frequently used sources of reliable atomic data is the Atomic Spectra Database (ASD) of the National Institute of Standards and Technology (NIST). In the current version of ASD (Kramida et al. 2021), the primary data source for the C II spectrum (wavelengths, line intensities, and energy levels) is Moore's tables (Moore 1970, 1993). These data are not in compliance with the present requirements of the ASD. In particular, they do not have uncertainties assigned to each individual value, many observed spectral lines are missing, and for many of them, only the Ritz wavelengths are given. The values of observed relative intensities are scarce, and they are given on a nonlinear scale that does not allow quantitative estimation of branching ratios for lines separated by large wavelength intervals. As mentioned above, Ritz wavelengths of carbon and its ions (including $C^+$) are widely used as secondary Ritz standards, especially in the VUV region. However, the absence of uncertainties on these quantities in the current ASD data sets impedes their use for this purpose. Thus, the primary aims of the present work are (i) to critically evaluate all available data on observed wavelengths; (ii) to derive a set of optimized energy levels and Ritz wavelengths with well-defined uncertainties; (iii) to derive an improved (i.e., more precise) value of the IE from the newly optimized energy levels; (iv) to provide a uniform description of observed line intensities; and (iv) to extend the list of critically evaluated A-values.

## 2. Evaluation of Observed Wavelengths

Methods of critical evaluation of atomic data have been described in our previous papers (Kramida 2013a, 2013b, 2013c; Haris et al. 2014; Haris & Kramida 2017). The same methods were used in the present work. Unlike the C I spectrum where numerous lines were precisely measured using FTS (Haris & Kramida 2017), only a few C II lines were found in FTS spectra. This is due to the limitation of FTSs, which require a stable and continuously operated light source over the period of the scan. Most laboratory carbon plasma sources are unstable or have pulsed operation due to the high ionization potential of carbon atoms (≈11.26 eV). The grating and prism instruments, on the contrary, are compatible with pulsed light sources allowing higher plasma temperatures. For this reason, we started our evaluation from Glad's (1953) comprehensive observations, made with grating and prism spectrometers, together with a few very accurate measurements of the lines connecting to the ground term made by Herzberg (1958). This armed us with rather accurate preliminary values of energy levels and Ritz wavelengths, especially in the VUV region. These preliminary values were used as references to evaluate the accuracy of other measurements and to include them in the level-optimization procedure, which was made with the LOPT code (Kramida 2011). This process was repeated iteratively until all measurements were evaluated. For each observed line, the most precisely measured wavelength was chosen for the final level optimization. An exception was made for a few

Table 1
Observed and Predicted Spectral Lines of C II

| $I_{obs}$[a] (arb. u.) | Char.[b] | $\lambda_{obs}$[c] (nm) | Unc.[d,e] (nm) | Lower Level[f,g] | | | Upper Level[f,g] | | | $\lambda_{Ritz}$[c] (nm) | Unc.[d] (nm) | $A(s^{-1})$ | Acc.[h,i] | TP Ref.[j] | Line_Ref.[j] | Weight[k] | Com.[l] |
|---|---|---|---|---|---|---|---|---|---|---|---|---|---|---|---|---|---|
| | a* | 4.2767 | 0.0004 | $2s^22p$ | $^2P°$ | 1/2 | $1s2s^22p^2$ | $^2S$ | 1/2 | 4.2766 | 0.0004 | 9.e+10 | D+ | M18LS | M18 | 0.333 | |
| | a* | 4.2767 | 0.0004 | $2s^22p$ | $^2P°$ | 3/2 | $1s2s^22p^2$ | $^2S$ | 1/2 | 4.2767 | 0.0004 | 1.7e+11 | D+ | M18LS | M18 | 0.667 | |
| ... | ... | ... | ... | ... | ... | ... | ... | ... | ... | ... | ... | ... | ... | ... | ... | ... | ... |
| 1100 | w* | 42.5326 | 0.0020 | $2s^22p$ | $^2P°$ | 1/2 | $2s2p(^3P°)6p$ | $^2S$ | 1/2 | 42.5250 | 0.0020 | 1.91e+08 | C+ | K01LS | E34,R01,K01 | 0.334 | C |
| 1100 | w* | 42.5326 | 0.0020 | $2s^22p$ | $^2P°$ | 3/2 | $2s2p(^3P°)6p$ | $^2S$ | 1/2 | 42.5364 | 0.0020 | 3.81e+08 | C+ | K01LS | E34,R01,K01 | 0.666 | C |
| ... | ... | ... | ... | ... | ... | ... | ... | ... | ... | ... | ... | ... | ... | ... | ... | ... | ... |
| 250000 | | 133.57079 | 0.00003 | $2s^22p$ | $^2P°$ | 3/2 | $2s2p^2$ | $^2D$ | 5/2 | 133.57077 | 0.00003 | 2.88e+08 | A | T00 | TW,H58,N21 | 1.000 | |
| ... | ... | ... | ... | ... | ... | ... | ... | ... | ... | ... | ... | ... | ... | ... | ... | ... | ... |
| 38 | h | 305.9091 | 0.0009 | $2s2p(^3P°)3d$ | $^4P°$ | 5/2 | $2s2p(^3P°)5f$ | $^4D$ | 7/2 | 305.9098 | 0.0007 | 6.74e+07 | C+ | TW | G53 | 1.000 | |
| ... | ... | ... | ... | ... | ... | ... | ... | ... | ... | ... | ... | ... | ... | ... | ... | ... | ... |
| 9 | | 400.1717 | 0.0007 | $2s2p(^3P°)3d$ | $^4D°$ | 5/2 | $2s2p(^3P°)4f$ | $^4G$ | 7/2 | 400.1715 | 0.0004 | 1.3e+06 | C | TW | G53 | 1.000 | N |
| ... | ... | ... | ... | ... | ... | ... | ... | ... | ... | ... | ... | ... | ... | ... | ... | ... | ... |
| 32 | * | 405.829 | 0.005 | $2s^24f$ | $^2F°$ | 5/2 | $2s^212g$ | $^2G$ | 7/2 | 405.8299 | 0.0022 | 1.6e+06 | C | TW | S03 | 0.429 | N |
| 32 | * | 405.829 | 0.005 | $2s^24f$ | $^2F°$ | 7/2 | $2s^212g$ | $^2G$ | 9/2 | 405.8328 | 0.0022 | 1.7e+06 | C | TW | S03 | 0.556 | N |
| 32 | * | 405.829 | 0.005 | $2s^24f$ | $^2F°$ | 7/2 | $2s^212g$ | $^2G$ | 7/2 | 405.8328 | 0.0022 | 6.e+04 | E | TW | S03 | 0.016 | N |
| ... | ... | ... | ... | ... | ... | ... | ... | ... | ... | ... | ... | ... | ... | ... | ... | ... | ... |
| 74 | * | 931.809 | 0.003 | $2s^25g$ | $^2G$ | 7/2 | $2s^28h$ | $^2H°$ | 9/2 | 931.806 | 0.003 | 3.69e+06 | B | N02 | S03 | 0.444 | N |
| 74 | * | 931.809 | 0.003 | $2s^25g$ | $^2G$ | 9/2 | $2s^28h$ | $^2H°$ | 9/2 | 931.812 | 0.003 | 8.4e+04 | D+ | N02 | S03 | 0.010 | N |
| 74 | * | 931.809 | 0.003 | $2s^25g$ | $^2G$ | 9/2 | $2s^28h$ | $^2H°$ | 11/2 | 931.812 | 0.003 | 3.78e+06 | B | N02 | S03 | 0.545 | N |
| ... | ... | ... | ... | ... | ... | ... | ... | ... | ... | ... | ... | ... | ... | ... | ... | ... | ... |
| | | 157740.92 | 0.05 | $2s^22p$ | $^2P°$ | 1/2 | $2s^22p$ | $^2P°$ | 3/2 | 157740.92 | 0.05 | 2.29e-06 | A | F04 | C86 | 1.000 | |

**Notes.** (A few columns are omitted in this condensed sample, but their footnotes (e, g, i) are retained for guidance regarding their form and content.)
[a] Averaged relative observed intensities of emission lines are given in arbitrary units on a uniform scale corresponding to Boltzmann populations in a plasma with an effective excitation temperature of 3.5 eV, corresponding to observations of Glad (1953), except for five intercombination lines near 232 nm, for which intensities are multiplied by an additional factor of 1000. For absorption lines (those having character "a"), intensities are roughly proportional to weighted absorption oscillator strengths. See Section 6 for details.
[b] Line character code: a–observed in absorption; bl–blended by other lines specified by elemental symbols and spectrum numbers, where possible; h–hazy; H–very hazy; i–identification uncertain; m–masked by other lines (no wavelength measurement was possible); l–shaded to long wavelength; p–perturbed by nearby lines; s–shaded to short wavelength; q–asymmetric; w–wide; *–intensity is shared by two or more lines.
[c] Observed and Ritz wavelengths are in standard air for 200 nm < λ < 2000 nm and in vacuum outside this range. Conversion between air and vacuum was made with the five-parameter formula from Peck & Reeder (1972).
[d] Assigned uncertainty of given observed wavelength and computed uncertainty of Ritz wavelength determined in the level optimization procedure. All uncertainties are on the 1σ level.
[e] Observed wavenumber (in vacuum) and its uncertainty. These two columns are omitted in this condensed sample of the table.
[f] Level designation from Table 2.
[g] Level energy value from Table 2. These two columns (for the lower and upper level) are omitted in this condensed sample of the table.
[h] Accuracy code of the A-value is given in Table 10 of Haris & Kramida (2017).
[i] Transition type: blank–electric–dipole (E1) transition; M1–magnetic–dipole transition; UT–forbidden transition of unknown type.
[j] Transition probability and line wavelength reference code: B26–Bowen & Ingram (1926) B27–Bowen (1927) B35–Boyce & Rieke (1935) B55–Bockasten (1955) B70–Burton & Ridgeley (1970) C86–Cooksy et al. (1986) C01–Curdt et al. (2001) C02–Corrégé & Hibbert (2002) D77–Doschek et al. (1977) E34–Edlén (1934) F04–Froese Fischer & Tachiev (2004) G53–Glad (1953) H58–Herzberg (1958) H84–Huber et al. (1984) K01–Kjeldsen et al. (2001) L21–Li et al. (2021) M18–Müller et al. (2018) N81–Nussbaumer & Storey (1981) N98–Nicolosi & Villoresi (1998); N02–Nahar (2002) N21–Nave & Clear (2021) P83–Penston et al. (1983) R01–Recanatini et al. (2001) S03–Sharpee et al. (2003) S13a–Sochi & Storey (2013) S13b–Storey & Sochi (2013) T99–Träbert et al. (1999) T00–Tachiev & Froese Fischer (2000) TW–this work; Y87–Yan et al. (1987) Y11–Young et al. (2011). "LS" after a reference code means that the multiplet values from the given reference have been decomposed into fine-structure components assuming a pure LS coupling; "n" after a reference code means that the transition probabilities given here were normalized to a scale different from that of the author(s) of that reference.
[k] Weight in the level optimization procedure.
[l] Comments: C–classification of the previous work(s) changed in the present work; N–newly identified line reported previously without classification; P–predicted line; S–single line that solely determines the upper energy level; ST–observed wavelength affected by a Stark shift.

(This table is available in its entirety in machine-readable form.)

Table 2
Energy Levels of C II

| Configuration[a] | Term[a] | Parity | J | Energy (cm$^{-1}$) | Unc.[b] (cm$^{-1}$) | Perc.[c] | Perc2[d] | Conf2[a] | Term2[a] | No. Lin[e] | $\Delta E_{o-c}$[f] (cm$^{-1}$) | Comm.[g] |
|---|---|---|---|---|---|---|---|---|---|---|---|---|
| $2s^2 2p$ | $^2P°$ | o | 1/2 | 0.000000 | 0.000000 | 96 | 3 | $2p^3$ | $^2P°$ | 83 | −1 | |
| $2s^2 2p$ | $^2P°$ | o | 3/2 | 63.395090 | 0.000022 | 96 | 3 | $2p^3$ | $^2P°$ | 116 | 0 | |
| $2s 2p^2$ | $^4P$ | e | 1/2 | 43,002.8 | 0.3 | 99 | | | | 41 | 2 | |
| $2s 2p^2$ | $^4P$ | e | 3/2 | 43,024.9 | 0.3 | 99 | | | | 66 | 2 | |
| $2s 2p^2$ | $^4P$ | e | 5/2 | 43,053.2 | 0.3 | 99 | | | | 58 | −5 | |
| $2s 2p^2$ | $^2D$ | e | 5/2 | 74,930.077 | 0.017 | 97 | | | | 13 | −1 | |
| $2s 2p^2$ | $^2D$ | e | 3/2 | 74,932.602 | 0.021 | 97 | | | | 12 | 2 | |
| $2s 2p^2$ | $^2S$ | e | 1/2 | 96,493.646 | 0.024 | 84 | 12 | $2s^2 3s$ | $^2S$ | 6 | 0 | |
| $2s 2p^2$ | $^2P$ | e | 1/2 | 110,624.18 | 0.04 | 96 | 2 | $2p^2(^3P)3s$ | $^2P$ | 10 | 0 | |
| $2s 2p^2$ | $^2P$ | e | 3/2 | 110,665.59 | 0.04 | 96 | 2 | $2p^2(^3P)3s$ | $^2P$ | 12 | −1 | |
| ... | ... | ... | ... | ... | ... | ... | ... | ... | ... | ... | ... | ... |
| $2s^2 4f$ | $^2F°$ | o | 5/2 | 168,978.24 | 0.05 | 94 | 6 | $2p^2(^1S)4f$ | $^2F°$ | 19 | 0 | FSC |
| $2s^2 4f$ | $^2F°$ | o | 7/2 | 168,978.42 | 0.04 | 94 | 6 | $2p^2(^1S)4f$ | $^2F°$ | 25 | 0 | FSC |
| ... | ... | ... | ... | ... | ... | ... | ... | ... | ... | ... | ... | ... |
| $2s^2 6h$ | $^2H°$ | o | 11/2 | 184,464.11 | 0.14 | 94 | 6 | $2p^2(^1S)6h$ | $^2H°$ | 0 | −2 | P,S |
| ... | ... | ... | ... | ... | ... | ... | ... | ... | ... | ... | ... | ... |
| $2s 2p(^3P°)4p$ | $^4D$ | e | 7/2 | 214,829.74 | 0.05 | 99 | | | | 7 | −121 | |
| ... | ... | ... | ... | ... | ... | ... | ... | ... | ... | ... | ... | ... |
| $1s 2s(^3S)\ 2p^3(^2P°)$ | $^4P°$ | o | 5/2 | 2,377,500 | 240 | 100 | | | | 2 | −4 | FSC |
| C III ($2s^2\ ^1S_0$) | Limit | | – | 196,663.31 | 0.10 | | | | | | | |

**Notes.**
[a] All level designations are in *LS* coupling. The complete $1s^2$ subshell is omitted from the configuration labels. Designations in the first two columns may correspond to an eigenvector component with the second largest percentage. They are chosen to uniquely define the level with the given *J* value and parity.
[b] Uncertainties of excitation energies from the ground level.
[c] Percentage of the configuration and term given in the first two columns in the eigenvector.
[d] Percentage of Conf2 and Term2 in the eigenvector.
[e] Number of observed lines determining the level in the level optimization procedure. Zero for unobserved levels.
[f] Difference of Energy from the value calculated in the parametric LSF procedure. Blank for unobserved levels or those that were not included in the LSF.
[g] Comments: FSC–In the level optimization, the fine-structure intervals within the term have been fixed at values calculated in the present work; FSI–In the level optimization, the fine-structure intervals within the term have been fixed at values determined in the present work by interpolation or extrapolation of differences between observed and calculated values of these intervals; P–The level value has been determined in this work by fitting the polarization formula; RDQ–The level value has been determined in this work by fitting the Ritz quantum defect expansion formula; S–The observed position of this level was affected by Stark shifts and thus was discarded.

(This table is available in its entirety in machine-readable form.)

dozen lines for which we used weighted average wavelengths from several observations (see Sections 2.2, 2.3, and 2.4 below). The list of all observed lines with wavelengths, observed intensities, and energy-level classifications is given in Table 1. This table also includes a large number of possibly observable predicted lines with their Ritz wavelengths. Standard uncertainties are specified for all observed and Ritz wavelengths, as well as for energy levels. In the main tables, they are given in separate columns; in the text and auxiliary tables, we specify them in parentheses after the value (in the units of the last decimal place of the value). The final level optimization yielded the energy levels listed in Table 2. The following subsections discuss the most important observations.

### 2.1. Observations of Glad

As mentioned above, Glad (1953) measured the C II spectrum in a wide range between 190 and 900 nm. Most of the measurements were made with a stigmatic 21 ft (6.4 m) concave grating spectrograph having a first-order dispersion of ≈0.5 nm mm$^{-1}$, operated in both first and second orders ($\lambda^{2nd} < 470$ nm). He also recorded some additional spectrograms on a large glass prism spectrograph having varying dispersion 0.1 to 0.77 nm mm$^{-1}$ for wavelengths $\lambda$ between 410 and 680 nm. Some of the weakest lines were recorded with a large quartz prism instrument having nonlinear dispersion 0.17 to 0.26 nm mm$^{-1}$ in the region between 220 nm and 250 nm. The source used was a condensed HC discharge with a pure graphite cathode tube embedded in a Pyrex bulb in vacuum. The carrier gas used for the excitation of the spectrum was a mixture of helium (99.9%) and neon (0.1%) at a pressure of 800 to 1600 Pa (i.e., between 6 Torr and 12 Torr). The source was operated in a pulse mode, with the help of a 10 μF capacitor charged to between 800 and 1800 V, with 20 to 30 sparks per second. The spectrograms were recorded on photographic plates of different types; the recording time varied from 1 minute to 90 minutes.

Glad (1953) used multiple wavelength standards (Ne I, Ar I, and Fe I– III) for different regions of wavelengths. The standard lines of argon and iron were obtained with the introduction of a small amount of these materials in their respective forms, gas and powder, into the source. Fe III lines were used for $\lambda$ between 190 and 220 nm, Fe I–II lines were used for $\lambda$ from 220 to 600 nm, and Ne I and/or Ar I lines were used for $\lambda >$ 600 nm. Most of Glad's final wavelengths were weighted averages of two or more measurements, but some of the faintest lines were measured on one plate only. He did not specify uncertainties for individual lines but gave a general statement that the uncertainty varies between 0.0005 and 0.01 nm depending on the spectral range and sharpness of the lines. To estimate the uncertainties for individual lines, we used the differences between the observed and Ritz wavenumbers tabulated by Glad. From those, we calculated the differences $\delta\lambda_{\mathrm{obs-Ritz}}$ between the corresponding observed and Ritz wavelengths. The entire line list was divided into several groups of lines depending on their intensity and character, as well as on the precision of the given wavelength. The root-mean-square values of $\delta\lambda_{\mathrm{obs-Ritz}}$ in each group were adopted as the wavelength uncertainties. They vary between 0.0007 and 0.05 nm (see Table 1). The final line list contains a total of 415 observed wavelengths from Glad (1953) with newly evaluated uncertainties. These lines participate in 507 transitions that were used in the final level optimization (see Section 3).

Most of Glad's observed wavelengths agree with the Ritz values within their combined uncertainties. However, Glad (1953) noted that some of the lines originating from highly excited states ($n \geqslant 6$) were significantly broadened and shifted by the Stark effect caused by a strong electric field in his light source. The most striking evidence of the presence of strong external fields in Glad's experiment is his observation of parity-forbidden transitions $2s^2 4f\ ^2F^\circ$–$2s^2 nf\ ^2H^\circ$ ($n = 6, 7$). He also noted that some lines were noticeably shifted on spectrograms taken at higher voltages. In his final line list, Glad (1953) included the least-shifted measurements. Nevertheless, the presence of residual Stark shifts cannot be excluded in some of the lines. Moreover, Bockasten (1955) has measured two lines of C II near 383.2 and 383.6 nm using a sliding spark as a light source and demonstrated that Glad's wavelengths are significantly shifted to the red. Wherever possible, we replaced the wavelengths of such lines with those from the measurements of Sharpee et al. (2003, 2004; see details in Sections 2.5 and 7). However, we could not use Bockasten (1955) measurements, since they are inconsistent with several other lines reported by Glad (1953) from the same upper levels, $2s2p(^3P^\circ)3p\ ^2D_{3/2,5/2}$. Thus, although these levels ($^2D_{3/2}$ and $^2D_{5/2}$ at 188,581.26 cm$^{-1}$ and 188,615.01 cm$^{-1}$, respectively) have uncertainties of 0.05 and 0.08 cm$^{-1}$ in Table 2, which is the measure of the internal consistency of Glad's measurements, they are shifted by as much as $-0.47(14)$ and $-0.71(14)$ cm$^{-1}$ from their field-free values. In Table 1, we have marked several observed lines that are known to have been affected by Stark shifts with a note "ST" (meaning "Stark") in the last column. However, all lines observed by Glad (1953) that have the characters q, l, s, h, or H may also possess Stark shifts of an unknown magnitude.

### 2.2. High-precision Measurements

The splitting in the ground term $2s^2 2p\ ^2P^\circ$ reported in Moore's tables (Moore 1970, 1993) was 63.42(10) cm$^{-1}$. The forbidden transition between the fine-structure levels of this term is in the far-infrared region. Although both the magnetic-dipole (M1) and electric-quadrupole (E2) transitions are possible between these two levels, the M1 transition dominates by far (Froese Fischer & Tachiev 2004). This transition can be observed only in an extremely low-density environment. Its first observation at $\lambda = 157.4(4)$ $\mu$m was made in the study of spectra of the Orion Nebula (M42) and the Flame Nebula (NGC 2024), both located in the Orion constellation (Russell et al. 1980). Several years later, Cooksy et al. (1986) made a very precise laboratory measurement of this transition using the laser magnetic resonance method. Their reported frequencies are 1900536.9(7) MHz and 1900545.8(11) MHz for $^{12}$C and $^{13}$C, respectively, which corresponds to the wavelengths of 157.74093(6) $\mu$m and 157.74019(9) $\mu$m. They also reported all three hyperfine (HF) components for the $^{13}$C$^+$ isotope, where the HF splitting arises due to the nonzero nuclear spin (see Section 4). Note that Cooksy et al. (1986) gave the uncertainties on the level of two standard deviations ($2\sigma$). Here, we divided them by two to reduce to the $1\sigma$ level used for all data in the present work.

As noted in the Introduction, Herzberg (1958) measured two groups of transitions, three lines of the $2s^2 2p\ ^2P^\circ$–$2s2p^2\ ^2D$ array at $\lambda \approx 133.5$ nm and two lines of the $2s2p^2\ ^2D$–$2s^2 3p\ ^2P^\circ$ array at $\lambda \approx 176.0$ nm, respectively. He used a HC lamp filled with a Ne buffer gas at a pressure of 133.3 Pa (1 Torr) and operated at a current of 250 mA. The spectrograms were taken on a 3 m vacuum spectrograph in the fifth order of diffraction. For the wavelength calibration, Herzberg used Ritz standards from the Mg II spectrum (Risberg 1955) originating from the material of the cathode. His stated wavelength uncertainty was 0.00007 nm for all lines. In addition to C II lines, Herzberg reported many precisely measured wavelengths of C I, N I, O I, and Ar II within the range from 91.9 to 197.4 nm. His stated uncertainty appears to be consistent with the more accurate Ritz wavelengths available for some of these spectra, e.g., C I lines near $\lambda \approx 132.9$ nm (Haris & Kramida 2017). To verify the accuracy of the C II wavelengths, we measured them on one of the SiC FT spectrograms used in the work on C I (Haris & Kramida 2017). Methods of measurement of line positions, widths, and intensities, as well as calibration of FT spectrograms have been explained in detail in our previous paper (Haris & Kramida 2017). The SiC spectrogram was calibrated with a global multiplicative scale factor $k_{\mathrm{eff}} = 1.95(6) \times 10^{-6}$ derived from two lines of Si II measured by Griesmann & Kling (2000). As follows from the uncertainty in the calibration factor, the systematic uncertainties of these C II lines are six to nine times smaller than the total uncertainty of Herzberg (1958). This enabled us to verify that there are no discernible systematic errors in Herzberg's measurements (see Table 3). The uncertainties of our FTS measurements are comparable to those of Herzberg (1958) or even somewhat greater due to low intensities of the lines. They are dominated by the statistical uncertainties.

Recently, Nave & Clear (2021) reported an independent measurement of the three lines near 133.5 nm on one spectrogram (#4 in their Table 1). We independently measured these lines, as well as two lines near 176 nm, on another spectrogram (#5 in Table 1 of Nave & Clear 2021). To calibrate the wavelength scale, Nave & Clear used Ar II lines, as well as their Ritz wavelengths in Ge II and some additional lines that were measured against Ar II standards in their other spectrograms. All their measurements can be traced to the Ar II reference wavenumbers (Whaling et al. 1995). They obtained $k_{\mathrm{eff}} = 1.90(5) \times 10^{-6}$ for the spectrogram #5 in fair agreement with our value derived from Si II lines.

G. Nave & C. Clear (2020, private communication) provided to us the list of all lines they have measured in the SiC spectrogram #4. In addition to the published data for the three lines near 133 nm, this list contained the measured wavelengths of the two lines near 176 nm. For all fine-structure lines near 133 and 176 nm, the uncertainties of the three available measurements (this work, Nave & Clear 2021, and Herzberg 1958) are dominated by statistical errors. We derived the final

Table 3
Comparison of Wavelengths from Grating and Fourier Instruments

| Transition | $\lambda_{\text{Grating}}$ (nm) | $\lambda_{\text{FTS}}$ (nm) | $\Delta\lambda_{G-F}$ (pm) | References[a] | Note on Calibration[b] |
|---|---|---|---|---|---|
| $2s^22p\ ^2P°–2s2p^2\ ^2D$ | | | | | |
| 1/2–3/2 | 133.45323(7) | 133.45326(8) | −0.03(11) | H58, N21 | Mg II (R55), Ar II |
|  |  | 133.45313(10) | 0.10(12) | TW$_{\text{SiC}}$ | Si II (G00) |
| 3/2–3/2 | 133.56625(7) | 133.5661(4) | 0.1(4) | H58, N21 | Mg II (R55), Ar II |
|  |  | 133.5663(4) | −0.1(4) | TW$_{\text{SiC}}$ | Si II (G00) |
| 3/2–5/2 | 133.57077(7) | 133.57082(4) | −0.05(8) | H58, N21 | Mg II (R55), Ar II |
|  |  | 133.57075(6) | 0.02(9) | TW$_{\text{SiC}}$ | Si II (G00) |
| $2s2p^2\ ^2D–2s^23p\ ^2P°$ | | | | | |
| 5/2–3/2 | 176.03954(7) | 176.03938(8) | 0.16(11) | H58, N20 | Mg II (R55), Ar II |
|  |  | 176.03954(5) | 0.00(9) | TW$_{\text{SiC}}$ | Si II (G00) |
| 3/2–3/2 | ... | ... | ... | ... | ... |
| 3/2–1/2 | 176.08191(7) | 176.08174(5) | 0.17(9) | H58, N20 | Mg II (R55), Ar II |
|  |  | 176.08195(15) | −0.04(17) | TW$_{\text{SiC}}$ | Si II (G00) |
| $2s2p^2\ ^2S–2s^23p\ ^2P°$ | | | | | |
| 1/2–3/2 | 283.6710(7) | 283.66924(24) | 1.8(7) | G53, H84 | Fe I–II (B29, D38) |
|  |  | 283.66988(13) | 1.1(7) | TW | C I (H17)+Ne I–II (S04,K06) |
| 1/2–1/2 | 283.7603(7) | 283.75886(16) | 1.4(7) | G53, H84 | Fe I–II (B29, D38) |
|  |  | 283.75955(26) | 0.7(7) | TW | C I (H17)+Ne I–II (S04,K06) |
| $2s^23s\ ^2S–2s^23p\ ^2P°$ | | | | | |
| 1/2–3/2 | 657.805(2) | 657.8022(4) | 2.8(20) | G53, H84 | Ne I (B50) |
|  | 657.8043(4) | 657.80481(8) | −0.5(4) | H16, TW | Th I (P83), C I+Ne I–II |
| 1/2–1/2 | 658.288(2) | 658.2847(4) | 3.3(20) | G53, H84 | Ne I (B50) |
|  | 658.2874(4) | 658.28764(12) | −0.2(4) | H16, TW | Th I (P83), C I+Ne I–II |

**Notes.**
[a] Reference code for the values in columns 2 and 3: G53–Glad (1953) H16–Hakalla et al. (2016) H58–Herzberg (1958) H84–Huber et al. (1984) N20–Additional data provided to us by Nave & Clear (2020) N21–Nave & Clear (2021) TW$_{\text{SiC}}$–SiC FT spectrogram used in the work on C I by Haris & Kramida (2017) TW–this work, from the 82R07 FT spectrum, possibly the same as used by H84 (see text).
[b] Descriptions of the wavelength standards: species and the reference code for the source of standard wavelengths used by observers quoted in column 5, except for H84, for which no calibration description was given by the authors. References to sources of standard wavelengths used in the present work for the two lines at the bottom are the same as for the previous two lines. Key to the references: B29–Burns & Walters (1929, 1931) B50–Burns et al. (1950) D38–Dobbie (1938) G00–Griesmann & Kling (2000) H17–Haris & Kramida (2017) K06–Kramida & Nave (2006) P83–Palmer & Engleman (1983) S04–Saloman & Sansonetti (2004).

experimental wavelengths as weighted means of the three directly measured ones. These mean values are given in Table 1.

In addition to the above VUV FT measurements, Huber et al. (1984) reported accurate wavenumbers of four lines corresponding to transitions from the $2s^23p\ ^2P°$ term down to the $^2S_{1/2}$ levels of the $2s2p^2$ and $2s^23s$ configurations with $\lambda$ near 283.7 nm and 658 nm, respectively. The spectrogram used by Huber et al. (1984) was taken on a 1 m ($f/55$ IR-visible-UV) FTS of Kitt Peak National Observatory (Brault 1978), Tucson, USA, with a neon–carbon HC lamp. There is a noticeable discrepancy between the measurements of Huber et al. (1984) and Glad (1953): the Huber et al. wavelengths near 283.7 nm are shifted by +1.6(5) pm from those of Glad (1953), while those near 658 nm are shifted by +3.1(14) pm (see Table 3). Huber et al. (1984) did not describe their wavelength-calibration procedure. Since their primary goal was to measure the branching ratios and not the wavelengths, we assume that their calibration scale factor could be in error. This is indicated by the large total uncertainties of their measured wavenumbers, which are apparently dominated by calibration uncertainties. For this reason, we decided to reanalyze the spectrograms used by Huber et al. (1984), which we retrieved from the Virtual Solar Observatory archive (Hill et al. 2004). Out of six total, the best spectrogram, 1982/06/26R0.007 (hereafter called 82R07), was recorded by M. C. Huber using a Fe/Ne/CO HC lamp with 109 Pa (0.82 Torr) of Ne and a trace of CO at a discharge current of 0.50 A. This spectrogram was taken at a resolution of 0.06 cm$^{-1}$ in the 7664 to 44,591 cm$^{-1}$ region. The spectrogram is enriched with many atomic and molecular features. However, lines of Ne II were strong, due to which some lines of the C II were expected to be present as well. We used lines of three intrinsic species (C I, Ne I, and Ne II) to check the constituency of the wavenumber scale of that spectrogram. For all species, only transitions from low-excited states (e.g., 3s–3p) were considered for the derivation of the global calibration factor. The calibration factors determined for each species together with the number of lines used and reference to the source of standard wavenumbers (given in parentheses) are as follows: $-1.124(14) \times 10^{-6}$ (11 lines of C I; Haris & Kramida 2017), $-1.199(12) \times 10^{-6}$ (13 lines of Ne I; Saloman & Sansonetti 2004), and $-1.045(16) \times 10^{-6}$ (17 lines of Ne II; Kramida & Nave 2006). The differences between the above values amount to a few standard deviations. This can be due to different spatial origin of emission from these species in the HC lamp leading to differences in the light paths through the FTS. To obtain the calibration factor for the C II lines, we took a straight arithmetic average of the three values given above, which is $-1.15(5) \times 10^{-6}$. Both pairs of lines at $\lambda \approx$ 284 and 658 nm have been remeasured, and our resulting values are given in Tables 1 and 3. These new results resolve the discrepancy between measurements of Glad (1953) and Huber et al. (1984) and lead to more accurate energies of the levels involved (see Table 2).

A recent study on the spectrum of $^{12}C^{17}O$ molecule carried out by Hakalla et al. (2016) reported wavenumbers of two C II

lines at $\lambda \approx 658$ nm (see Table 3). The spectrum was recorded on a 2 m Ebert plane-grating spectrograph of instrumental resolution of 0.09–0.23 cm$^{-1}$ in the region of 15,180–18,400 cm$^{-1}$. For calibration purposes, they used the known lines of Th I (Palmer & Engleman 1983) from a simultaneously recorded spectrum of an auxiliary water-cooled thorium HC lamp. Relative positions of molecular lines were measured very accurately, to $(2.5–8) \times 10^{-4}$ cm$^{-1}$. However, an absolute calibration uncertainty of 0.002 cm$^{-1}$ was determined from the residuals of the Th I calibration. The two C II lines were reported to be at 15,197.891 cm$^{-1}$ and 15,186.739 cm$^{-1}$. No uncertainties were specified for these measurements. These lines were observed with significantly greater intensities and line widths ($W \approx 0.5$ cm$^{-1}$) in comparison to the molecular CO lines ($W \approx 0.15$ cm$^{-1}$). By comparing the measured wavenumbers of the C II and other impurity lines (of hydrogen and deuterium), we roughly estimate the uncertainties of these two C II wavenumbers as ±0.010 cm$^{-1}$.

As seen from Table 3, our new FTS measurements agree well with the FTS measurements of G. Nave & C. Clear (2020, private communication), Nave & Clear (2021). They also agree with the best grating-spectroscopy data but provide a much greater precision than the latter.

### 2.3. Grating Measurements of Vacuum Ultraviolet Lines

In C II, unlike C I, observed VUV transitions terminate not only on the levels of the ground term but also on many other low-excited levels. It was already known from earlier publications (Edlén 1934; Glad 1953; Herzberg 1958) that accurate VUV wavelengths can be calculated using the Ritz combination principle from the energy levels established by lines with longer wavelengths. We used this method to derive preliminary reference wavelengths that were subsequently used to evaluate the uncertainties of VUV observations. Most of the available observed wavelengths were reported by Edlén (1934). However, a few dozen lines were measured with a comparable or better accuracy by other authors (Bowen & Ingram 1926; Bowen 1927; Boyce & Rieke 1935; Curdt et al. 2001). Unlike the latter two studies, the wavelengths reported by Bowen & Ingram (1926) and by Bowen (1927) were found to contain significant systematic errors, which we removed by recalibrating the reported wavelengths against the more accurate Ritz values available in the NIST ASD (Kramida et al. 2021; for C I, III, and IV, as well as for other species) together with the newly determined C II Ritz wavelengths. For the wavelengths reported by Edlén (1934) and by Boyce & Rieke (1935), we did not find any discernible systematic errors. Unlike Edlén (1934), Boyce & Rieke explicitly stated the uncertainty for each wavelength. From our comparisons, we found that those given uncertainties should be combined in quadrature with an additional 0.0003 nm to bring their given wavelengths into statistical agreement with the reference values. For Edlén's data, we separately assessed the uncertainties of wavelengths given with two or three digits after the decimal point (in angstroms), with different character (indicated along with the observed intensity), and for singly and multiply classified lines. Uncertainties vary between 0.0008 and 0.0024 nm among these categories of lines.

Curdt et al. (2001) gave a list of wavelengths measured in solar spectra recorded with the Solar Ultraviolet Measurements of Emitted Radiation (SUMER) VUV spectrograph on board Solar and Heliospheric Observatory (SOHO) spacelab. This list contains a few dozen C II lines. Curdt et al. (2001) stated that their wavelengths should be accurate to typically 10 mÅ (0.001 nm). However, all wavelengths were given with only two digits after the decimal point (in angstroms), which introduced significant rounding errors. The reciprocal dispersion of the SUMER instrument is about 43 mÅ pixel$^{-1}$, i.e., 0.0043 nm pixel$^{-1}$ (Curdt et al. 1997), and the average full line width at half maximum on the detector is 5.5 pixels (Feldman et al. 1997). Thus, the FWHM of fully resolved isolated lines is about 0.24 Å (0.024 nm) on average. Curdt et al. (2001) have listed 1393 lines between 67 and 161 nm. Many of them were severely blended. Curdt et al. (2001) extensively applied line-profile decomposition to extract the wavelengths of individual components of the blends. When all lines contributing to a blended feature were known, their relative positions on the wavelength scale could be extracted with a good precision of about 0.005 Å (0.0005 nm). However, this did not always work that well. For the calibration of the wavelength scale, Curdt et al. (2001) used reference wavelengths taken mostly from the line list of Kelly (1987). In general, the use of this line list as a source of reference wavelengths is discouraged, since it does not provide measurement uncertainties. Some of the lines listed by Kelly (1987) are now known to contain large errors. Nevertheless, when several hundred lines are used in a wide wavelength range, such as in Curdt et al. (2001), these errors are expected to average out. We checked the wavelengths listed by Curdt et al. (2001) against Ritz wavelengths of 428 lines of C I, O II, N II, Fe II, and Fe III taken from the NIST ASD (Kramida et al. 2021) and, indeed, found that the average errors of the measurements of Curdt et al. (2001) are about 0.01 Å (0.001 nm), in agreement with their own assessment. However, many of their listed wavelengths deviate from the much more precise reference values by more than 0.02 Å (0.002 nm), up to 0.14 Å (0.014 nm), which implies either wrong identifications or the presence of undetected severe blending. Presence of such unexpected blending cannot be excluded for any of the lines in the list of Curdt et al. (2001). Thus, to be on the somewhat safer side, we assigned uncertainties of 0.002 nm to all their measured wavelengths. Three C II multiplets observed as single blended lines in the old laboratory spectra have been partially resolved in the solar line list of Curdt et al. (2001). Another solar atlas by Parenti et al. (2005) from the same SOHO/SUMER instrument contains 24 lines of C II with estimated uncertainties ranging between 0.0022 and 0.0051 nm. However, none of these wavelengths were found to be more accurate than other available values. Thus, they were not used in the present work.

The VUV measurements competing in precision with each other are listed in Table 4. In the final level optimization, we used a weighted average for each of these lines (given in the last column of Table 4). The line at 132.39 nm corresponds to the $2s2p^2\ ^2D–2p^3\ ^2D°$ multiplet. It was completely unresolved in the two laboratory measurements (Edlén 1934; Boyce & Rieke 1935), but partially resolved into two components by Curdt et al. (2001). Each of these two components is in turn a blend of two fine-structure transitions (see Table 1). We used the data of Table 4 to locate the center of gravity of the multiplet and then applied the splitting observed by Curdt et al. (2001) to derive the wavelengths given in Table 1. The partially resolved blends at 80.666 and 80.685 nm were treated similarly.

Table 4
Observed VUV Wavelengths (nm) of Competing Precision that Were Averaged in the Final Line List

| Edlén (1934) | Bowen & Ingram (1926) [a] | Bowen (1927) [a] | Boyce & Rieke (1935) | Curdt et al. (2001) 1st order | Curdt et al. (2001) 2nd order | Mean |
|---|---|---|---|---|---|---|
| 59.4808(17) |  | 59.4792(10) |  |  |  | 59.4796(10) |
| 59.5032(17) |  | 59.5022(10) |  |  |  | 59.5025(9) |
| 68.7059(10) |  | 68.7055(5) |  | 68.705(2) |  | 68.7056(5) |
| 79.9664(10) |  |  |  | 79.966(2) |  | 79.9663(9) |
| 79.9947(15) |  |  |  | 79.994(2) |  | 79.9944(12) |
| 80.6555(15) |  |  |  | 80.6559(15)[b] |  | 80.6557(11) |
| 80.6684(15) |  |  |  | 80.668(2) |  | 80.6682(12) |
| 80.6846(15) |  |  |  | 80.6846(14)[b] |  | 80.6846(10) |
| 80.9682(15) |  |  |  | 80.968(2) |  | 80.9681(12) |
| 68.7355(15) |  | 68.7353(5) |  | 68.735(2) | 68.7345(10) | 68.7352(5) |
| 85.8094(8) |  | 85.8089(5) | 85.8091(4) | 85.804(2) |  | 85.8090(3) |
| 85.8561(8) |  | 85.8562(5) | 85.8559(4) | 85.853(2) |  | 85.8560(3) |
| 90.3609(8) |  | 90.3621(5) | 90.3614(9) | 90.359(2) |  | 90.3616(5) |
| 90.3950(8) |  | 90.3961(5) | 90.3952(9) | 90.399(2) |  | 90.3958(5) |
| 90.4134(8) |  | 90.4134(5) | 90.4144(4) | 90.414(2) |  | 90.4139(4) |
| 90.4468(8) |  | 90.4473(5) | 90.4482(4) | 90.446(2) |  | 90.4477(4) |
| 94.5981(10) |  |  |  | 94.599(2) |  | 94.5983(9) |
| 94.6208(10) |  |  |  | 94.619(2) |  | 94.6204(10) |
| 100.9854(8) | 100.9856(9) | 100.9870(5) | 100.9862(4) | 100.985(2) |  | 100.9863(4) |
| 101.0074(8) | 101.0088(9) | 101.0090(5) | 101.0092(4) | 101.003(2) |  | 101.0088(4) |
| 101.0369(8) | 101.0376(9) | 101.0382(5) | 101.0374(4) | 101.037(2) |  | 101.0376(4) |
| 103.6330(8) | 103.6340(5) | 103.6336(5) | 103.6332(4) | 103.634(2) |  | 103.6335(3) |
| 103.7017(8) | 103.7020(5) | 103.7021(5) | 103.7020(4) | 103.700(2) |  | 103.7020(3) |
| 106.5883(10) |  |  | 106.5895(4) | 106.586(2) |  | 106.5892(4) |
| 106.6121(10) |  |  | 106.6138(4) | 106.613(2) |  | 106.6135(5) |
| 109.2740(10) |  |  |  | 109.273(2) |  | 109.2738(9) |
| 113.8936(10) |  |  |  | 113.894(2) |  | 113.8937(9) |
| 113.9330(10) |  |  | 113.9343(9) | 113.938(2) |  | 113.9341(9) |
| 114.1630(10) |  | 114.1609(10) | 114.1623(4) | 114.168(2) |  | 114.1622(4) |
| 114.1746(10) |  |  | 114.1745(6) |  |  | 114.1745(5) |
| 132.3916(8) |  |  | 132.3940(6) | 132.3934(14)[b] |  | 132.3932(8) |

**Notes.**
[a] Wavelengths of Bowen & Ingram (1926) and of Bowen (1927) have been corrected by removing systematic errors (see text).
[b] Weighted mean of two resolved components with weights proportional to calculated intensities.

### 2.4. Intercombination Lines from the Lowest Quartet Term $2s2p^2\,{}^4P$

The $2s^22p\,{}^2P°$–$2s2p^2\,{}^4P$ intercombination transitions are difficult to observe under usual laboratory conditions due to their low radiative decay rates. These transitions (located between 232 and 233 nm) have diagnostic applications in a variety of laboratory and astrophysical plasmas (Smith et al. 1999; Young et al. 2011). The energies of the $2s2p^2\,{}^4P_{1/2,3/2,5/2}$ levels were established in Moore's tables (Moore 1970, 1993) via VUV transitions from them to the upper quartet levels observed in laboratory light sources. These levels, in turn, were determined via the $2s2p3d$-$2s2pnf$ ($n = 4$–$6$) intercombination transitions observed by Glad (1953), which are enabled by a strong departure of the $2s2pnf$ configurations from pure $LS$ coupling. However, the five transitions from $2s2p^2\,{}^4P$ to the levels of the ground term were observed and measured only in solar off-limb and limb emission spectra (Burton & Ridgeley 1970; Doschek et al. 1977) and in emission of the RR Tel Nebula (Penston et al. 1983; Young et al. 2011). All these measurements had comparable uncertainties of ≈0.005 nm. For each transition, we adopted a weighted average of wavelengths from these three studies. Prior to averaging, the wavelengths reported in old studies (Burton & Ridgeley 1970; Doschek et al. 1977; Penston et al. 1983) had been slightly corrected by using improved wavelength standards taken from the NIST ASD (Kramida et al. 2021). The mean wavelengths have significantly smaller uncertainties, ≈0.0027 nm on average, which lead to a more accurate determination of the $2s2p^2\,{}^4P$ levels. It is worth mentioning that the radiative rates ($A$-values) of these five intercombination transitions were precisely measured by Träbert et al. (1999) using a storage ring setup.

### 2.5. The Spectrum of the IC 418 Nebula Spectra

Here we will discuss observations of C II lines in emission spectra of nebulae. Many of these observations include spectral lines that are difficult or impossible to observe in laboratory experiments and are used in the present work to construct a complete set of experimental data for C II.

Spectra of astrophysical objects often contain features that are impossible or very difficult to observe in laboratory light sources. This is due to unique physical conditions such as very low particle densities or strong photoexcitation by a nearby star. On the other hand, there is a problem of overlapping lines from multiple species that are hard to disentangle. Spectra of stars and nebulae can also be affected by Doppler shifts, which can have a significant spread even within the same object. This complicates the analysis of line shapes and can cause errors in interpretation of observed features. With this in mind, we reviewed several astrophysical studies that reported observation of C II lines.

Lines of C II have been observed in the emission spectra of many nebulae (Kaler 1976; Sharpee et al. 2003, 2004; Zhang et al. 2005; Otsuka et al. 2010; Fang & Liu 2011; García-Rojas et al. 2012). In particular, the catalog of Kaler (1976) gives wavelengths and intensities of lines observed in over 600 objects, most containing C II lines. Most of those old observations are of a too low precision to be considered. The advent of cross-dispersed echelle spectrometers in the 1990 s led to a significant increase in the resolving power available to astronomers. The most extensive and best-quality data were reported by Sharpee et al. (2003, 2004). They recorded high-resolution and high signal-to-noise ratio (S/N) spectra of the planetary nebula IC 418 (also known as Spirograph Nebula) on a ground-based Blanco telescope at Cerro Tololo Inter-American Observatory, Chile, with a cross-dispersed echelle spectrometer. Two different setups called blue and red were used to record spectra in the 350–595 nm and 509–986.5 nm ranges. These two setups employed echelle and cross-dispersing gratings of different resolutions, as well as different optics and filters. The red configuration was used in two different settings in the 509–742.5 nm and 735–986.5 nm regions. The narrowest emission lines of IC 418 were found to have the full width at half maximum of 15 km s$^{-1}$, which corresponds to 0.03 nm at $\lambda = 600$ nm. To maintain the calibration of the spectrometer during the five nights of observations, Sharpee et al. (2003, 2004) recorded the reference spectrum of a Th-Ar HC lamp once per hour. Several exposures of different durations were coadded to reduce the S/N. They stated that the measured wavelengths are accurate to about 1 km s$^{-1}$, which corresponds to 0.002 nm at $\lambda = 600$ nm. More details about measurement procedures and uncertainties can be found in Sharpee's thesis (Sharpee 2003). The thesis explains that the above estimate describes the systematic uncertainty stemming from the calibration of the spectrograph and stitching of exposures covering the adjacent spectral regions.

The line list of Sharpee et al. (2003) contains 806 distinct wavelengths, 691 of which are assigned to 1005 transitions in 37 atomic spectra (H I, He I, C I–III, N I–III, O I–III, Ne I–III, Na I, Mg I–II, Al II, Si II–III, P II, S II–III, Cl II–III, Ar II–IV, Ca I–II, Fe I–III, Co II, Ni I–III). To identify the observed lines, these authors used a computer-aided emission line-identification software package EMILI (Sharpee et al. 2003). As described in their subsequent paper (Sharpee et al. 2004), they found that different spectra (and some different groups of lines within the same spectrum) originate from different spatial regions of the nebula and are caused by different level-population processes. The main contribution to the wavelength uncertainty is due to the nonuniformity of velocity distribution within the nebula. Different velocities along the line of sight cause different Doppler shifts, which is the main cause of the observed line widths and shifts. The analysis of wavelength uncertainties described by Sharpee (2003) was based on statistical comparisons of measured wavelengths with reference values taken from van Hoof's Atomic Line List database (van Hoof 1999). These reference wavelengths were computed from energy levels taken mainly from an old version of ASD (v1.1., 1997) and are not supplemented by uncertainties. The current version of ASD (Kramida et al. 2021) provides more accurate energy levels and Ritz wavelengths, many of which are supplemented by critically evaluated uncertainties. We reanalyzed the wavelengths measured by Sharpee et al. (2003) by comparing them with the new set of reference values (Kramida et al. 2021). This analysis accounted for the S/N and line widths reported by Sharpee et al. (2003) and was done separately for each part of the spectrum of each species originating from the same spatial region of the nebula.

We found that the hydrogen lines measured by Sharpee et al. (2003) are systematically shifted from the rest-frame reference values. This shift corresponds to a mean velocity difference of 0.36(9) km s$^{-1}$. Our estimates of total uncertainties for the hydrogen wavelengths reported by Sharpee et al. (2003) range from 0.0005 to 0.01 nm, depending on S/N and line width. They are smaller than those used by Sharpee (2003) in the input for the line-identification code (for seven lines given in Sharpee's thesis) by a factor of seven on average.

For C II lines, we do not find any statistically significant systematic difference of the mean velocity from the value given by Sharpee (2003), 61.3(9) km s$^{-1}$. More precisely, the weighted mean velocity difference is found to be 0.10(34) km s$^{-1}$, which is statistically consistent with a zero value. Similarly, for He I lines, we find the mean velocity difference of $-0.01(20)$ km s$^{-1}$, also consistent with zero.

Sharpee et al. (2003) found that lines originating from the outer cold and fast-expanding shell of the nebula appear as well-resolved doublets due to large differential expansion velocities along the line of sight. These lines are due to species with low IE and include forbidden lines within the ground configurations, e.g., [N I] and [C I]. However, lines of spectra with higher IEs appear to come from higher-temperature central regions of the nebula. These lines (including those of C II) appear as single Gaussian-shaped peaks. Another observation of Sharpee et al. (2003, 2004) is that, within one species, the intensities of lines from the levels populated by different processes are correlated with concentrations of different ions. For example, C II lines from levels populated by collisional excitation or photoabsorption from the ground state are correlated with a concentration of $C^+$, while those from levels populated by recombination (either radiative or dielectronic) are correlated with a concentration of $C^{2+}$. Lines originating from hotter species (with large IEs) are observed to have narrower profiles. Deviations of line profiles from average for a given species indicate that more than one population mechanism is at play for some lines.

As usual in astrophysical spectra, many observed lines are blends of two or more transitions from either the same or different species. Compared to other astrophysical objects, IC 418 has an advantage of containing mostly light species. The heaviest identified species is nickel (Ni) with a few observed forbidden lines of the first three spectra. On the other hand, due to a very low electron density, the spectra of light species are excited to very high principal quantum numbers $n$. For example, the Balmer and Paschen series of H I are observed up to $n = 42$ and $n = 43$, respectively. The highly excited series in H I, He I, N I–II, and O I–II produce a very large number of lines, many of which blend with each other and with C II lines. Sharpee et al. (2004) identified 83 lines of C II. Using our refined Ritz wavelengths and series formulas (see further sections), we identified 14 more lines of this spectrum. Out of total 97 lines associated with C II, based on our analysis of observed line widths, intensities, and wavelengths, seven lines are deemed to be masked by other species, 12 other lines are blended with other species, and for another 13 lines, the identification is questionable. Out of 65 remaining C II lines, five were found to be perturbed by lines of other species in the sense that these other species contribute a significant fraction of the observed intensity.

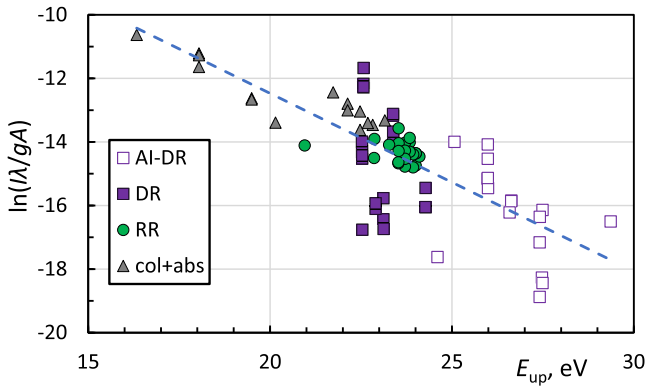

**Figure 1.** Boltzmann plot for lines of C II observed by Sharpee et al. (2003, 2004) in IC 418. $I$ is the observed intensity as reported by Sharpee et al. (2003), $\lambda$ is the observed wavelength in Å, $gA$ is the weighted radiative transition probability in s$^{-1}$, and $E_{up}$ is the upper level energy in eV. Empty and solid squares denote lines from autoionizing and nonautoionizing $2s2p(^3P^\circ)nl$ levels, respectively, which are populated mainly by dielectronic recombination (DR). Circles and triangles denote lines from levels populated by radiative recombination (RR) and by electronic collisions and photoabsorption (col+abs). The dotted line is a linear fit to all data points. Its slope is $-0.56\,\text{eV}^{-1}$, corresponding to an effective excitation temperature of 1.8 eV.

Since the current knowledge of high-$n$ energy levels and transitions between them is very scarce for N I–II, O I–II, and other species observed in IC 418, there is no guarantee that any of the observed lines are free from blending with a currently unknown transition. Thus, we assumed that any deviation of the observed line width from the average for a given species may be due to such blending. Another possible reason, as mentioned above, is a contribution of an unknown peculiar process to the population of the upper level, which changes the spatial distribution of the source of the emitted light compared to the average one for a given ionic or atomic species. Blending by an unresolved line of comparable intensity may shift the fitted wavelength of the peak. The possible shift can be estimated as half the difference of the observed line width from the nominal (average) one.

The problem of the unknown spatial origin of lines enabled by different level-population processes can be investigated with the help of Boltzmann plots. Figure 1 shows a Boltzmann plot for the observed C II lines of different origin. In this plot, the empty and solid squares denote the lines believed to be enabled by dielectronic recombination originating from autoionizing and nonautoionizing levels (AI-DR and DR lines, respectively). The upper levels of all these lines are of the type $2s2p(^3P^\circ)nl$ ($n = 3\text{–}6, l = s, p, d, f$). Circles denote the lines enabled mainly by radiative recombination (RR lines). The upper levels of these lines belong to the $2s^2nl$ ($n = 4\text{–}14$, $l = f, g, h$) configurations. Triangles denote the lines with upper levels belonging to singly excited configurations $2s^2nl$ ($n = 3\text{–}7$) with low orbital angular momentum $l = s, p, d$. These levels are populated mainly by two processes: collisional excitation and photoabsorption, both from the ground state, followed by radiative cascades. In assigning different population processes to different levels, we followed the considerations of Storey & Sochi (2013) for physical conditions in planetary nebulae, as well as those of Sharpee et al. (2004).

Sharpee et al. (2004) argued that population of the upper levels of all recombination lines (both DR and RR) are in equilibrium with the ground state of $C^{2+}$, while the upper levels of all other lines are linked to the population of the ground state of $C^+$.

However, Figure 1 shows that observed intensities ($I$) of all lines are described reasonably well by a model assuming all excited levels to be in local thermodynamic equilibrium (LTE). It strongly suggests that $C^+$ and $C^{2+}$ coexist in the same spatial regions of the nebula, and their relative abundances roughly obey the Saha distribution. Sharpee et al. (2004) have determined the electron density $N_e$ and electron temperature $T_e$ to be about $10^5\,\text{cm}^{-3}$ and 0.8 eV, respectively (see their Table 2). It is almost obvious that at these conditions, supplemented by intense radiation from the central star, the plasma is not in full thermodynamic equilibrium. Indeed, the effective excitation temperature determined from the slope of the dotted line in Figure 1 is about 1.8 eV. Saha-Boltzmann modeling with the LIBS interface to ASD (Kramida et al. 2021) shows that, at $N_e = 10^5\,\text{cm}^{-3}$ and $T_e = 1.8\,\text{eV}$, the carbon emission from a hydrogen-carbon mixture would be dominated by lines of C III, while none of these lines are detected in the spectrum of IC 418. (Note: in the line list of Sharpee et al. (2003), the only C III assignment of a doubly classified feature at 465.1559 nm to the $2s3s\ ^3S_1 - 2s3p\ ^3P^\circ_0$ transition is incorrect, since this transition must be accompanied by the other two much stronger components of the same multiplet, which were not observed in IC 418.)

The only distinction between the different groups of lines shown in Figure 1 is the significantly greater scatter of the data points of the DR (and AI-DR) lines compared to other lines. It may mean that intensities of the DR lines are more sensitive to various local factors, such as the density, temperature, and proximity of their place of origin to the central star. Their interpretation requires more careful treatment with spatially resolved spectroscopy and collisional-radiative modeling, while intensities of all other lines are fairly well described by a simple model with Boltzmann populations of levels. The average deviation of non-DR lines from the straight fitted line in Figure 1 is by a factor of 1.7, while it is by a factor of 4 for DR lines. These observations allowed us to make definite conclusions about relative contributions of C II to several blended lines observed by Sharpee et al. (2003). In particular, if the C II intensities predicted from the Boltzmann plot shown in Figure 1 were more than an order of magnitude smaller than observed, we concluded that the C II contribution to the observed blend is negligibly small. This is the case, for example, for the lines observed at 498.736 nm (N II), 639.2496 nm ([Fe II]), 716.0559 (He I), 821.6298 nm (N I), 823.4505 nm (Mg II), 828.5688 nm (He I), and 843.3723 nm ([Cl III]). The line at 672.4226 nm was assigned to two C II transitions. One of them, $2s2p(^3P^\circ)4d\ ^4D^\circ_{3/2} - 2s2p(^3P^\circ)6p\ ^4D_{1/2}$, is predicted to be 440 times weaker than the observed feature, so we discarded this assignment.

Most of the previously unidentified lines listed by Sharpee et al. (2003) that we now identify with C II (see Table 1) are due to transitions from highly excited $2s^2nl$ levels ($n \leqslant 14$, $l = f, g, h$). These identifications were supported by the accurate predictions of the level values using a polarization formula (see Section 8).

## 2.6. Photoabsorption Lines from the Lowest $^2P^\circ$ and $^4P$ Terms

Photoabsorption spectra of C II involving excitation of the K and L shells were studied by several groups of authors using different techniques. The K-shell photoabsorption spectrum in the extreme ultraviolet (XUV) region from 3.8 to 4.4 nm was first reported by Jannitti et al. (1993) who used the dual laser-produced plasmas (LPP) setup. In this setup, one LPP provides a source of strong backlighting continuous radiation, and the second LPP supplies the studied absorbing medium. Unfortunately, theoretical interpretation of Jannitti et al. (1993) was based solely on matching the observed wavelengths with theoretical predictions without paying attention to the population of the lower states of those matching transitions. As a result, most of their suggested identifications are incorrect. Furthermore, the deconvolution technique they applied to their spectrograms produced many narrow peaks that are nothing more than noise.

Recently, Müller et al. (2018) precisely measured the energies of six lines corresponding to $1s \to 2p$ excitation resonances associated with absorption from the ground term $^2P^\circ$ and from the lowest metastable $^4P$ term. These lines are in the narrow region between 4.27 and 4.32 nm. Each of these lines is a blend of two or more fine-structure transitions. In the experiment of Müller et al. (2018), an isotopically pure beam of $^{14}$C$^+$ ions was illuminated by monochromatic synchrotron radiation (bandwidth of 12 meV at 288 eV energy), and the ions produced by photoionization were counted by a sensitive detector. The high S/Ns achieved in this experiment allowed extraction of not only the resonance energies and widths but also the absorption oscillator strengths and partial rates for single-, double-, and triple-electron Auger decay.

It should be noted that the energies of the $1s2s^22p^2$ $^4P$, $^2P$, $^2D$, and $^2S$ terms were measured earlier by Rødbro et al. (1979) using Auger electron spectroscopy. Each of these energies can be determined from several observed peaks corresponding to different final states of C$^{2+}$. These individual measurements agree well with each other, except for one assignment of the peak at 249.3(7) eV to the $1s2s^22p^2$ $^2S \to 1s^22s2p$ $^1P^\circ$ transition, which must be discarded. From the data of Müller et al. (2018), this transition should be located at 252.84(3) eV. According to our calculations, its intensity should be smaller than the peak at 251.5(5) eV by a factor of 150, which means that the observed peak at 252.6(6) eV is dominated by some other transitions, perhaps the two Auger transitions of C III assigned to it by Rødbro et al. (1979). The weighted mean energies of doublet terms determined from observations of Rødbro et al. (1979) agree with the data of Müller et al. (2018) within their uncertainties of about 0.4 eV. However, a detailed comparison of measured peak energies with those precisely defined by the energies measured by Müller et al. (2018) in C II, as well as with other precise reference data on C III and C IV (Scully et al. 2005; Yerokhin et al. 2017), shows that the energies reported by Rødbro et al. (1979) for carbon are systematically too high by 0.13(3) eV. Removing this calibration error and correcting a few erroneous line assignments allowed us to extract the energy of the C II $1s2s^22p^2$ $^4P$ level, 2,295,800(800) cm$^{-1}$, from the data of Rødbro et al. (1979) with improved precision.

Nicolosi & Villoresi (1998) observed spectral absorption features for several $2s^2nd$ $^2D$ ($n \leqslant 12$) and $2s2p(^3P^\circ)n\ell$ $^{2,\,4}L$ ($n \leqslant 8$, $\ell = s,\ p,\ d$) manifolds using a dual-laser-plasma-technique. Absorption was initiated from the ground term $2s^22p$ $^2P$ and from the lowest metastable term $2s2p^2$ $^4P$. Recanatini et al. (2001) later revised these measurements with twice-greater accuracy. Both the backlighting and analyzed plasmas were created by splitting a pulsed beam of a Nd:YAG laser, part of which was focused onto a tungsten target to produce the backlighting plasma, and the remaining part was focused on a graphite target to create an absorbing plasma column. A normal incidence spectrograph equipped with a toroidal grating with 3600 lines mm$^{-1}$ groves having a linear inverse dispersion of 0.275 nm mm$^{-1}$ in the first order of diffraction was used to record the spectrum. The detection system was a charge-coupled device (CCD) that was able to record a spectral range of ≈3.4 nm in one setting. Several exposures were combined to cover the spectral range from 40 to 70 nm. Both C II and C III lines were observed, and some known lines of carbon served as calibration standards. For C II, many photoionization resonances were observed by both Nicolosi & Villoresi (1998) and Recanatini et al. (2001). Due to the limited resolution of the CCD and broadening caused by the Doppler and Stark effects, no fine-structure components were resolved. The line identifications were largely based on a simple quantum-defect extrapolation along the Rydberg series and on several previously reported calculations of other authors.

The absorption spectrum of C II from the $2s^22p$ $^2P$ and $2s2p^2$ $^4P$ terms was also investigated by Kjeldsen et al. (1999, 2001). These authors used an ion-photon interaction setup with a beam of C$^+$ ions illuminated by a SRS. Unlike the dual laser plasma conditions, where approximately 15% of C$^+$ ions were excited to the metastable $2s2p^2$ $^4P$ term (Recanatini et al. 2001), the fraction of metastable excitations was much smaller (≈5%) in the work of Kjeldsen et al. (2001). As a result, they observed only three weak peaks corresponding to excitation from the $2s2p^2$ $^4P$ term, while the remaining 35 measured peaks were associated with absorption from the ground $2s^22p$ $^2P$ term. The identifications were assisted by their own extensive Cowan code calculations and by comparisons with the R-matrix calculations of Nahar (1995). Kjeldsen et al. (1999, 2001) reported observed spectral lines in the range between 25 and 51 nm and stated that the absolute uncertainty of the measured peak energies is 0.015 eV, which corresponds to an uncertainty of 0.008–0.31 nm in wavelength. They mentioned that the relative precision (i.e., precision of measurements of separations between the peaks) is better for well-resolved resonances. However, their tabulated data give no indication of relative precision except for the number of significant digits in the given energies. To better understand the quality of recorded data and to obtain numerical estimates of observed intensities, we digitized the tracings of the recorded photoionization cross section given in the figures of Kjeldsen et al. (2001). Several peaks were found to be too weak and noisy to allow reliable fitting with a combination of Fano and Gaussian profiles, as it was done by Kjeldsen et al. (2001). We discarded a few of such unreliable measurements and increased the uncertainties for the measured wavelengths of a few peaks.

Measurements of Recanatini et al. (2001) and Kjeldsen et al. (2001) overlap in the region from 41 to 44 nm. The wavelengths measured by both teams agree well within the quoted uncertainties. For all lines observed in both these experiments, we selected the measurements of Recanatini et al. (2001), as they are more accurate. Assignments of observed peaks to theoretical transitions agreed in both works for most of the commonly observed lines, with only a few exceptions. We reviewed these assignments with the help of our parametric least-squares fitting (LSF) calculations of energy levels and line intensities. These Cowan code calculations are described in Section 6. The results of our analysis are reflected in Table 1. We revised some of the original classifications and identified a few lines previously reported as unclassified.

## 3. Energy Levels and Their Optimization

To determine the energy levels that best fit all observed wavelengths, we have used the computer code LOPT (Kramida 2011). Its use was described in detail in several previous papers, e.g., Kramida (2013b, 2013c) and Haris et al. (2014). Since the Ritz wavelengths were used for the evaluation of measurement uncertainties, levels were optimized several times. The first optimization was made with only the wavelengths of Glad (1953) in the input file (see Section 2.1), including those VUV lines from Edlén (1934) that he used to determine the energy levels. Then we included a few lines accurately measured by Cooksy et al. (1986), Herzberg (1958), and four lines of C II from the National Solar Observatory's FTS 82R07 spectrogram (see Section 2.2). These additions were supplemented by inclusion of the critically evaluated VUV wavelengths, which provide an accurate connection of excited terms to the ground level (see Section 2.3), and by the intercombination lines described in Section 2.4. At this stage, we were able to accurately predict wavelengths of many previously unreported transitions and search for lines observed near these wavelengths in the literature. We found many such lines in the line list of Sharpee et al. (2003, 2004; see Section 2.5).

As noted in the previous section, many observed lines, especially in the VUV and XUV regions, are unresolved blends of fine-structure transitions between narrow manifolds. Even in the visible range, some of the multiplets observed either in the laboratory (Glad 1953) or in the nebula spectra (Sharpee et al. 2003) are partially or completely unresolved, e.g., the $2s^2 4d$ $^2D$–$2s^2 7p$ $^2P°$ transition at 536.858 nm and the $2s^2 4d$ $^2D$–$2s^2 6f$ $^2F°$ transition at 615.143 nm. However, in order to account for all resolved fine-structure transitions and to make presentation of the energy structure internally consistent, it was necessary to explicitly present the unresolved blends as separate transitions with the same wavelength in the input file for level optimization. For several Rydberg series, such as $2s^2 nf$ $^2F°$, $2s^2 ng$ $^2G$, and $2s^2 nh$ $^2H°$, the fine-structure intervals within the unresolved terms were fixed at values from our parametric LSF calculations. For other series, such as $2s2p(^3P°)ns$ $^{2,4}S°$, $2s2p(^3P°)np$ $^{2,4}P$, and $2s2p(^3P°)nd$ $^{2,4}D°$, the small differences between the observed fine-structure intervals and those from our LSF calculations were extrapolated along the series. In total, 37 fine-structure intervals were fixed at such calculated or extrapolated values in our level-optimization procedure.

The final results of our level optimization are presented in Tables 1 and 2. For energy levels in Table 2, one uncertainty value is given, which represents the uncertainty of excitation energy from the ground level. It was determined as a maximum of two values, $D_1$ and $D_2$, returned by the LOPT code. $D_1$ provides a measure of minimum relative uncertainty of separations from other levels (see Kramida 2011 for the exact definition of this quantity), while $D_2$ gives an estimate of uncertainty of excitation energy from the ground level. That estimate ignores deviations of observed wavelengths from the Ritz values, some of which may stem from line blending or other quasi-random systematic effects. These deviations are accounted for in $D_1$. In many atomic spectra, including C II, uncertainties of separations between some of the excited levels are much smaller than those of separations from the ground level, as the measurements of wavenumbers of long-wavelength transitions are usually more precise than those at short wavelengths. In such cases, rounding the energy levels according to the precision of excitation energy would lead to adversely large errors in Ritz wavenumbers. In our level-optimization procedure for C II, for many levels, $D_1$ was significantly smaller than $D_2$, which explains the necessity for keeping additional significant digits in the energy value. Thus, rounding of the level values in Table 2 was based on the minimum of $D_1$ and $D_2$. For transitions that are not included in Table 1, Ritz wavenumbers can be calculated from the levels given in Table 2, and their uncertainties can be roughly estimated as the sum in quadrature of the uncertainty values from Table 2 for the two levels of the transition.

## 4. Isotope Shifts

As mentioned in our previous work on C I (Haris & Kramida 2017), natural carbon samples are composed of two stable isotopes, $^{12}$C (98.94%, nuclear spin $I = 0$) and $^{13}$C (1.06%, $I = 1/2$), and traces of unstable $^{14}$C (<$10^{-4}$%, $I = 0$, half-life $\tau_{1/2} = 5700(30)$ yr) (Meija et al. 2016; Audi et al. 2017). For C II, isotope shifts (IS) have been measured for only a few spectral lines (Burnett 1950, 1957; Cooksy et al. 1986; Haridass & Huber 1994). However, recent developments in atomic theory enabled accurate calculations of IS (Berengut et al. 2006; Korol & Kozlov 2007; Jönsson et al. 2010; Bubin & Adamowicz 2011; Hornyák et al. 2020; S. Bubin & L. Adamowicz 2021, private communication). We critically evaluated all these theoretical data based on their agreement with the experiment and with other theoretical data. The results of this evaluation are listed in Table 5 for transition wavenumbers in $^{12}$C$^+$, $^{13}$C$^+$, and $^{14}$C$^+$. This table includes transitions from the ground level to all levels with known IS, which makes a separate level table unnecessary.

As noted above, Cooksy et al. (1986) made absolute frequency measurements for the $2s^2 2p$ $^2P°$ $J = 1/2$–$3/2$ transition in both $^{12}$C$^+$ and $^{13}$C$^+$. In addition to that, they analyzed the isotope shift $\Delta E(^{13}\text{C}-^{12}\text{C})$ and obtained a value of 8.9(8) MHz for this parameter (we remind again that the $2\sigma$ uncertainties given by Cooksy et al. (1986) have been divided by 2 here).

Haridass & Huber (1994) measured the wavenumbers of the three fine-structure components of the $2s^2 2p$ $^2P°$–$2s2p^2$ $^2D$ multiplet near $\lambda = 133$ nm. Their uncertainties were about 0.10 cm$^{-1}$. To derive the ($^{13}$C–$^{12}$C) IS, they used old reference data for natural carbon from Kaufman & Edlén (1974) in place of $^{12}$C$^+$ values. Instead, we used much more accurate Ritz wavenumbers obtained in this work for natural carbon and used the formula

$$\text{IS}(^{13}\text{C} - ^{12}\text{C}) = (\sigma_{13} - \sigma_{\text{nat}})/c_{12}, \qquad (1)$$

where $\sigma_{13}$ and $\sigma_{\text{nat}}$ are the wavenumbers for $^{13}$C and for natural carbon, respectively, and $c_{12} = 0.9894(10)$ is the average relative abundance of $^{12}$C in natural carbon samples (Meija et al. 2016). Similarly, if the isotope shift between $^{13}$C and $^{12}$C is known, the wavenumbers of transitions in these isotopes can easily be calculated from the optimized Ritz wavenumbers

**Table 5**
Transition Wavelengths and Isotope Shifts for C II

| Lower Level | Upper Level | $\lambda$[a] (C II)nm | $K_{MS}$ (inp)[b] (GHz u) | Ref.[c] | $K_{MS}$ (opt)[d] (GHz u) | IS(13−12)cm$^{-1}$ | IS(14−12)cm$^{-1}$ | $\sigma$($^{12}$C II)cm$^{-1}$ | $\sigma$($^{13}$C II)cm$^{-1}$ | $\sigma$($^{14}$C II)cm$^{-1}$ |
|---|---|---|---|---|---|---|---|---|---|---|
| $2s^22p\ ^2P°_{1/2}$ | $2s^27p\ ^2P°_{3/2}$ | 53.54858(7) | −506(35) | B21 | −506(35) | 0.109(8) | 0.201(14) | 186746.33(25) | 186746.44(25) | 186746.53(25) |
| $2s^22p\ ^2P°_{1/2}$ | $2s^27p\ ^2P°_{1/2}$ | 53.54885(13) | −506(35) | B21 | −506(35) | 0.109(8) | 0.201(14) | 186745.36(46) | 186745.47(46) | 186745.56(46) |
| $2s^22p\ ^2P°_{1/2}$ | $2s^26p\ ^2P°_{3/2}$ | 54.64686(4) | −537(35) | B21 | −537(35) | 0.115(8) | 0.214(14) | 182993.12(15) | 182993.23(15) | 182993.33(15) |
| $2s^22p\ ^2P°_{1/2}$ | $2s^26p\ ^2P°_{1/2}$ | 54.64693(5) | −537(35) | B21 | −537(35) | 0.115(8) | 0.214(14) | 182992.91(18) | 182993.02(18) | 182993.12(18) |
| $2s^22p\ ^2P°_{1/2}$ | $2s^26s\ ^2S_{1/2}$ | 55.168083(24) | −362(35) | B21 | −363(35) | 0.078(8) | 0.144(14) | 181264.229(80) | 181264.306(80) | 181264.374(81) |
| $2s^22p\ ^2P°_{3/2}$ | $2s^26s\ ^2S_{1/2}$ | 55.187384(24) | | | −362(35) | 0.078(8) | 0.144(14) | 181200.829(80) | 181200.906(80) | 181200.973(81) |
| $2s^22p\ ^2P°_{1/2}$ | $2s2p(^3P°)3s\ ^2P°_{3/2}$ | 56.245020(16) | −2998(35) | B21 | −2999(94) | 0.644(20) | 1.19(4) | 177793.513(50) | 177794.150(54) | 177794.706(62) |
| $2s^22p\ ^2P°_{1/2}$ | $2s2p(^3P°)3s\ ^2P°_{1/2}$ | 56.251021(19) | −2998(35) | B21 | −2998(94) | 0.643(20) | 1.19(4) | 177774.543(60) | 177775.180(63) | 177775.736(71) |
| $2s^22p\ ^2P°_{1/2}$ | $2s^25p\ ^2P°_{3/2}$ | 57.046777(13) | −2629(35) | B21 | −2632(57) | 0.565(12) | 1.047(23) | 175294.734(40) | 175295.293(42) | 175295.781(46) |
| $2s^22p\ ^2P°_{1/2}$ | $2s^25p\ ^2P°_{1/2}$ | 57.049149(13) | −2629(35) | B21 | −2632(57) | 0.565(12) | 1.047(23) | 175287.354(40) | 175287.913(42) | 175288.401(46) |
| $2s^22p\ ^2P°_{1/2}$ | $2s^25s\ ^2S_{1/2}$ | 57.68745(4) | −225(35) | B21 | −224(35) | 0.048(8) | 0.089(14) | 173347.94(13) | 173347.99(13) | 173348.03(13) |
| $2s^22p\ ^2P°_{3/2}$ | $2s^25s\ ^2S_{1/2}$ | 57.70855(3) | | | −223(35) | 0.048(8) | 0.089(14) | 173284.54(10) | 173284.59(10) | 173284.63(10) |
| $2s^22p\ ^2P°_{1/2}$ | $2p^3\ ^2P°_{3/2}$ | 59.259870(11) | −5180(53) | J10,B21 | −5181(52) | 1.112(11) | 2.061(21) | 168748.248(30) | 168749.348(32) | 168750.309(36) |
| $2s^22p\ ^2P°_{1/2}$ | $2p^3\ ^2P°_{1/2}$ | 59.266457(12) | −5176(50) | J10,B21 | −5177(49) | 1.111(11) | 2.060(19) | 168729.493(35) | 168730.592(37) | 168731.553(40) |
| $2s^22p\ ^2P°_{1/2}$ | $2s2p(^3P°)3s\ ^4P°_{5/2}$ | 59.867432(18) | −3475(55) | J10 | −3475(55) | 0.746(12) | 1.383(22) | 167035.719(50) | 167036.457(51) | 167037.102(55) |
| $2s^22p\ ^2P°_{1/2}$ | $2s2p(^3P°)3s\ ^4P°_{3/2}$ | 59.883559(18) | −3477(55) | J10 | −3477(55) | 0.746(12) | 1.383(22) | 166990.734(50) | 166991.472(51) | 166992.117(55) |
| $2s^22p\ ^2P°_{1/2}$ | $2s2p(^3P°)3s\ ^4P°_{1/2}$ | 59.892018(18) | −3478(55) | J10 | −3478(55) | 0.746(12) | 1.384(22) | 166967.149(51) | 166967.888(52) | 166968.533(56) |
| $2s^22p\ ^2P°_{1/2}$ | $2s^24p\ ^2P°_{3/2}$ | 61.529162(12) | −525(31) | J10,B21 | −525(31) | 0.113(7) | 0.209(12) | 162524.559(31) | 162524.670(32) | 162524.768(33) |
| $2s^22p\ ^2P°_{1/2}$ | $2s^24p\ ^2P°_{1/2}$ | 61.531684(13) | −526(31) | J10,B21 | −526(31) | 0.113(7) | 0.209(12) | 162517.897(35) | 162518.008(36) | 162518.106(37) |
| $2s^22p\ ^2P°_{1/2}$ | $2s^24s\ ^2S_{1/2}$ | 63.599456(12) | 48(30) | J10,B21 | 49(30) | −0.011(6) | −0.019(12) | 157234.050(30) | 157234.040(31) | 157234.031(32) |
| $2s^22p\ ^2P°_{3/2}$ | $2s^24s\ ^2S_{1/2}$ | 63.625109(12) | | | 50(30) | −0.011(6) | −0.020(12) | 157170.650(30) | 157170.639(31) | 157170.630(32) |
| $2s^22p\ ^2P°_{1/2}$ | $2p^3\ ^2D°_{3/2}$ | 66.459888(22) | −7960(110) | J10 | −7891(73) | 1.693(16) | 3.14(3) | 150466.682(50) | 150468.357(52) | 150469.821(58) |
| $2s^22p\ ^2P°_{1/2}$ | $2p^3\ ^2D°_{5/2}$ | 66.46215(3) | −7960(110) | J10 | −7835(65) | 1.681(14) | 3.12(3) | 150461.562(70) | 150463.226(71) | 150464.679(75) |
| $2s^22p\ ^2P°_{1/2}$ | $2s^23d\ ^2D_{5/2}$ | 68.704582(14) | 344(55) | J10 | 344(55) | −0.074(12) | −0.137(22) | 145550.701(30) | 145550.628(32) | 145550.564(37) |
| $2s^22p\ ^2P°_{1/2}$ | $2s^23d\ ^2D_{3/2}$ | 68.705266(13) | 344(55) | J10 | 344(55) | −0.074(12) | −0.137(22) | 145549.251(30) | 145549.178(32) | 145549.114(37) |
| $2s^22p\ ^2P°_{3/2}$ | $2s^23d\ ^2D_{5/2}$ | 68.734520(15) | | | 346(55) | −0.074(12) | −0.137(22) | 145487.301(30) | 145487.227(32) | 145487.163(37) |
| $2s^22p\ ^2P°_{3/2}$ | $2s^23d\ ^2D_{3/2}$ | 68.735204(13) | | | 346(55) | −0.074(12) | −0.137(22) | 145485.861(30) | 145485.787(32) | 145485.723(37) |
| $2s^22p\ ^2P°_{1/2}$ | $2p^3\ ^4S°_{3/2}$ | 70.40952(15) | −7126(80) | J10 | −7126(80) | 1.529(17) | 2.84(3) | 142026.22(30) | 142027.74(30) | 142029.06(30) |
| $2s^22p\ ^2P°_{1/2}$ | $2s^23p\ ^2P°_{3/2}$ | 75.909667(11) | −45(15) | B06,K07,J10, B11,B21 | −62(11) | 0.0133(24) | 0.025(5) | 131735.528(19) | 131735.541(19) | 131735.552(20) |
| $2s^22p\ ^2P°_{1/2}$ | $2s^23p\ ^2P°_{1/2}$ | 75.916090(11) | −45(15) | B06,K07,J10, B11,B21 | −31(12) | 0.007(3) | 0.012(5) | 131724.382(19) | 131724.389(19) | 131724.394(20) |
| $2s2p^2\ ^4P_{3/2}$ | $2s2p(^3P°)3s\ ^4P°_{5/2}$ | 80.63812(20) | | | 194(57) | −0.042(12) | −0.077(23) | 124010.80(30) | 124010.76(30) | 124010.72(30) |
| $2s2p^2\ ^4P_{1/2}$ | $2s2p(^3P°)3s\ ^4P°_{3/2}$ | 80.65300(18) | | | 193(57) | −0.041(12) | −0.077(23) | 123987.90(30) | 123987.86(30) | 123987.82(30) |
| $2s2p^2\ ^4P_{5/2}$ | $2s2p(^3P°)3s\ ^4P°_{5/2}$ | 80.65653(21) | | | 194(57) | −0.042(12) | −0.077(23) | 123982.50(30) | 123982.46(30) | 123982.42(30) |
| $2s2p^2\ ^4P_{3/2}$ | $2s2p(^3P°)3s\ ^4P°_{3/2}$ | 80.66738(20) | | | 192(57) | −0.041(12) | −0.077(23) | 123965.80(30) | 123965.76(30) | 123965.72(30) |
| $2s2p^2\ ^4P_{1/2}$ | $2s2p(^3P°)3s\ ^4P°_{1/2}$ | 80.66835(18) | | | 192(57) | −0.041(12) | −0.076(23) | 123964.40(30) | 123964.36(30) | 123964.32(30) |
| $2s2p^2\ ^4P_{3/2}$ | $2s2p(^3P°)3s\ ^4P°_{1/2}$ | 80.68273(20) | | | 191(57) | −0.041(12) | −0.076(23) | 123942.30(30) | 123942.26(30) | 123942.22(30) |
| $2s2p^2\ ^4P_{5/2}$ | $2s2p(^3P°)3s\ ^4P°_{3/2}$ | 80.68580(21) | | | 192(57) | −0.041(12) | −0.076(23) | 123937.50(30) | 123937.46(30) | 123937.42(30) |
| $2s2p^2\ ^4P_{1/2}$ | $2s^24p\ ^2P°_{3/2}$ | 83.66677(19) | | | 3145(34) | −0.675(7) | −1.251(14) | 119521.81(30) | 119521.14(30) | 119520.56(30) |
| $2s2p^2\ ^4P_{1/2}$ | $2s^24p\ ^2P°_{1/2}$ | 83.67144(19) | | | 3144(34) | −0.675(7) | −1.251(14) | 119515.11(30) | 119514.44(30) | 119513.86(30) |
| $2s2p^2\ ^4P_{3/2}$ | $2s^24p\ ^2P°_{3/2}$ | 83.68225(21) | | | 3144(34) | −0.675(7) | −1.251(14) | 119499.71(30) | 119499.04(30) | 119498.46(30) |
| $2s2p^2\ ^4P_{3/2}$ | $2s^24p\ ^2P°_{1/2}$ | 83.68691(21) | | | 3143(34) | −0.675(7) | −1.251(14) | 119493.01(30) | 119492.34(30) | 119491.76(30) |
| $2s2p^2\ ^4P_{5/2}$ | $2s^24p\ ^2P°_{3/2}$ | 83.70207(22) | | | 3144(34) | −0.675(7) | −1.251(14) | 119471.41(30) | 119470.74(30) | 119470.16(30) |
| $2s^22p\ ^2P°_{1/2}$ | $2s^23s\ ^2S_{1/2}$ | 85.809180(14) | 294(30) | J10,B21 | 292(30) | −0.063(6) | −0.116(12) | 116537.649(19) | 116537.587(20) | 116537.532(22) |
| $2s^22p\ ^2P°_{3/2}$ | $2s^23s\ ^2S_{1/2}$ | 85.855884(14) | | | 293(30) | −0.063(6) | −0.117(12) | 116474.254(19) | 116474.191(20) | 116474.137(22) |
| $2s^22p\ ^2P°_{1/2}$ | $2s2p^2\ ^2P_{3/2}$ | 90.36232(3) | −3311(28) | B07,J96,J10 | −3319(27) | 0.712(6) | 1.321(11) | 110665.582(40) | 110666.287(40) | 110666.903(41) |

Table 5
(Continued)

| Lower Level | Upper Level | λ[a] (C II)nm | $K_{MS}$ (inp)[b] (GHz u) | Ref.[c] | $K_{MS}$ (opt)[d] (GHz u) | IS(13−12)cm$^{-1}$ | IS(14−12)cm$^{-1}$ | σ($^{12}$C II)cm$^{-1}$ | σ($^{13}$C II)cm$^{-1}$ | σ($^{14}$C II)cm$^{-1}$ |
|---|---|---|---|---|---|---|---|---|---|---|
| $2s^22p$ $^2P°_{1/2}$ | $2s2p^2$ $^2P_{1/2}$ | 90.39615(3) | −3312(28) | B07,J96,J10 | −3317(28) | 0.712(6) | 1.320(11) | 110624.172(40) | 110624.877(40) | 110625.492(41) |
| $2s^22p$ $^2P°_{3/2}$ | $2s2p^2$ $^2P_{3/2}$ | 90.41412(3) | | | −3318(27) | 0.712(6) | 1.320(11) | 110602.182(40) | 110602.887(40) | 110603.502(41) |
| $2s^22p$ $^2P°_{3/2}$ | $2s2p^2$ $^2P_{1/2}$ | 90.44798(3) | | | −3315(28) | 0.711(6) | 1.319(11) | 110560.772(40) | 110561.476(40) | 110562.091(41) |
| $2s2p^2$ $^4P_{1/2}$ | $2p^3$ $^4S°_{3/2}$ | 100.9862(3) | | | −3456(81) | 0.742(17) | 1.37(3) | 99023.39(30) | 99024.13(30) | 99024.77(30) |
| $2s2p^2$ $^4P_{3/2}$ | $2p^3$ $^4S°_{3/2}$ | 101.0087(3) | | | −3457(81) | 0.742(17) | 1.38(3) | 99001.29(30) | 99002.03(30) | 99002.67(30) |
| $2s2p^2$ $^4P_{5/2}$ | $2p^3$ $^4S°_{3/2}$ | 101.0376(3) | | | −3457(81) | 0.742(17) | 1.38(3) | 98972.99(30) | 98973.73(30) | 98974.37(30) |
| $2s^22p$ $^2P°_{1/2}$ | $2s2p^2$ $^2S_{1/2}$ | 103.63377(3) | −2930(22) | B07,K07,J96,J10,B11,B21 | −2921(11) | 0.6269(24) | 1.162(5) | 96493.639(24) | 96494.260(24) | 96494.802(24) |
| $2s^22p$ $^2P°_{3/2}$ | $2s2p^2$ $^2S_{1/2}$ | 103.70190(3) | | | −2920(11) | 0.6266(24) | 1.162(5) | 96430.244(24) | 96430.864(24) | 96431.406(24) |
| $2s2p^2$ $^2D_{5/2}$ | $2p^3$ $^2P°_{3/2}$ | 106.58915(4) | | | −1284(55) | 0.275(12) | 0.511(22) | 93818.177(30) | 93818.450(32) | 93818.688(37) |
| $2s2p^2$ $^2D_{3/2}$ | $2p^3$ $^2P°_{3/2}$ | 106.59201(4) | | | −1286(55) | 0.276(12) | 0.512(22) | 93815.657(30) | 93815.930(32) | 93816.169(37) |
| $2s2p^2$ $^2D_{3/2}$ | $2p^3$ $^2P°_{1/2}$ | 106.61333(4) | | | −1282(52) | 0.275(11) | 0.510(21) | 93796.897(30) | 93797.169(32) | 93797.407(36) |
| $2s2p^2$ $^2D_{5/2}$ | $2s2p(^3P°)3s$ $^4P°_{5/2}$ | 108.57097(6) | | | 422(58) | −0.091(12) | −0.168(23) | 92105.651(50) | 92105.561(51) | 92105.483(55) |
| $2s2p^2$ $^2D_{3/2}$ | $2s2p(^3P°)3s$ $^4P°_{5/2}$ | 108.57395(6) | | | 420(58) | −0.090(12) | −0.167(23) | 92103.131(50) | 92103.042(51) | 92102.964(55) |
| $2s2p^2$ $^2D_{5/2}$ | $2s2p(^3P°)3s$ $^4P°_{3/2}$ | 108.62403(6) | | | 420(58) | −0.090(12) | −0.167(23) | 92060.661(50) | 92060.572(51) | 92060.494(55) |
| $2s2p^2$ $^2D_{3/2}$ | $2s2p(^3P°)3s$ $^4P°_{3/2}$ | 108.62700(6) | | | 418(58) | −0.090(12) | −0.166(23) | 92058.141(50) | 92058.052(51) | 92057.975(55) |
| $2s2p^2$ $^2D_{3/2}$ | $2s2p(^3P°)3s$ $^4P°_{1/2}$ | 108.65484(6) | | | 417(58) | −0.090(12) | −0.166(23) | 92034.561(50) | 92034.472(51) | 92034.395(55) |
| $2s2p^2$ $^2S_{1/2}$ | $2s^27p$ $^2P°_{3/2}$ | 110.80003(23) | | | 2415(37) | −0.518(8) | −0.961(15) | 90252.69(19) | 90252.17(19) | 90251.72(19) |
| $2s2p^2$ $^2S_{1/2}$ | $2s^27p$ $^2P°_{1/2}$ | 110.8012(3) | | | 2415(37) | −0.518(8) | −0.961(15) | 90251.72(24) | 90251.20(24) | 90250.75(24) |
| $2s2p^2$ $^4P_{1/2}$ | $2s^23p$ $^2P°_{3/2}$ | 112.6980(4) | | | 3608(19) | −0.774(4) | −1.436(7) | 88732.71(30) | 88731.94(30) | 88731.27(30) |
| $2s2p^2$ $^4P_{1/2}$ | $2s^23p$ $^2P°_{1/2}$ | 112.7121(4) | | | 3639(19) | −0.781(4) | −1.448(8) | 88721.61(30) | 88720.84(30) | 88720.16(30) |
| $2s2p^2$ $^4P_{3/2}$ | $2s^23p$ $^2P°_{3/2}$ | 112.7261(4) | | | 3608(19) | −0.774(4) | −1.435(7) | 88710.61(30) | 88709.84(30) | 88709.17(30) |
| $2s2p^2$ $^4P_{3/2}$ | $2s^23p$ $^2P°_{1/2}$ | 112.7402(4) | | | 3638(19) | −0.781(4) | −1.448(8) | 88699.51(30) | 88698.74(30) | 88698.06(30) |
| $2s2p^2$ $^4P_{5/2}$ | $2s^23p$ $^2P°_{3/2}$ | 112.7620(4) | | | 3607(19) | −0.774(4) | −1.435(7) | 88682.31(30) | 88681.54(30) | 88680.87(30) |
| $2s2p^2$ $^2D_{5/2}$ | $2s^24p$ $^2P°_{3/2}$ | 114.16244(4) | | | 3372(35) | −0.724(8) | −1.342(14) | 87594.488(30) | 87593.772(31) | 87593.146(33) |
| $2s2p^2$ $^2D_{3/2}$ | $2s^24p$ $^2P°_{3/2}$ | 114.16573(4) | | | 3370(35) | −0.723(8) | −1.341(14) | 87591.968(30) | 87591.252(31) | 87590.627(33) |
| $2s2p^2$ $^2D_{3/2}$ | $2s^24p$ $^2P°_{1/2}$ | 114.17442(4) | | | 3369(35) | −0.723(8) | −1.340(14) | 87585.308(30) | 87584.592(31) | 87583.967(33) |
| $2s2p^2$ $^2S_{1/2}$ | $2s^26p$ $^2P°_{3/2}$ | 115.60764(20) | | | 2384(37) | −0.512(8) | −0.948(15) | 86499.48(15) | 86498.97(15) | 86498.53(15) |
| $2s2p^2$ $^2S_{1/2}$ | $2s^26p$ $^2P°_{1/2}$ | 115.6079(2) | | | 2384(37) | −0.512(8) | −0.948(15) | 86499.27(18) | 86498.76(18) | 86498.32(18) |
| $2s2p^2$ $^2S_{1/2}$ | $2s2p(^3P°)3s$ $^2P°_{3/2}$ | 123.00142(6) | | | −78(95) | 0.017(20) | 0.03(4) | 81299.870(40) | 81299.886(45) | 81299.901(55) |
| $2s2p^2$ $^2S_{1/2}$ | $2s2p(^3P°)3s$ $^2P°_{1/2}$ | 123.03013(8) | | | −77(95) | 0.017(20) | 0.03(4) | 81280.900(60) | 81280.916(63) | 81280.931(71) |
| $2s2p^2$ $^2S_{1/2}$ | $2s^25p$ $^2P°_{3/2}$ | 126.90179(5) | | | 289(58) | −0.062(12) | −0.115(23) | 78801.091(30) | 78801.029(32) | 78800.976(38) |
| $2s2p^2$ $^2S_{1/2}$ | $2s^25p$ $^2P°_{1/2}$ | 126.91368(6) | | | 289(58) | −0.062(12) | −0.115(23) | 78793.711(40) | 78793.649(42) | 78793.596(46) |
| $2s2p^2$ $^2D_{5/2}$ | $2p^3$ $^2D°_{3/2}$ | 132.38611(9) | | | −3993(75) | 0.857(16) | 1.59(3) | 75536.611(50) | 75537.459(52) | 75538.200(58) |
| $2s2p^2$ $^2D_{3/2}$ | $2p^3$ $^2D°_{3/2}$ | 132.39054(9) | | | −3995(75) | 0.857(16) | 1.59(3) | 75534.091(50) | 75534.939(52) | 75535.680(58) |
| $2s2p^2$ $^2D_{5/2}$ | $2p^3$ $^2D°_{5/2}$ | 132.39509(11) | | | −3938(67) | 0.845(14) | 1.57(3) | 75531.491(60) | 75532.327(62) | 75533.058(66) |
| $2s2p^2$ $^2D_{3/2}$ | $2p^3$ $^2D°_{5/2}$ | 132.39951(11) | | | −3940(67) | 0.845(14) | 1.57(3) | 75528.971(60) | 75529.808(62) | 75530.539(66) |
| $2s^22p$ $^2P°_{1/2}$ | $2s2p^2$ $^2D_{3/2}$ | 133.45326(3) | −3896(17) | B07,K07,J96,J10 | −3895(17) | 0.836(4) | 1.550(7) | 74932.593(19) | 74933.420(19) | 74934.143(20) |
| $2s^22p$ $^2P°_{1/2}$ | $2s2p^2$ $^2D_{5/2}$ | 133.45776(3) | −3897(17) | B06,K07,J96,J10 | −3897(17) | 0.836(4) | 1.551(7) | 74930.068(16) | 74930.896(16) | 74931.619(17) |
| $2s^22p$ $^2P°_{3/2}$ | $2s2p^2$ $^2D_{3/2}$ | 133.56626(3) | −3600(330) | H94 | −3894(17) | 0.836(4) | 1.549(7) | 74869.198(19) | 74870.025(19) | 74870.747(20) |
| $2s2p^2$ $^2P°_{3/2}$ | $2s2p^2$ $^2D_{5/2}$ | 133.57077(3) | −4180(470) | H94 | −3896(17) | 0.836(4) | 1.550(7) | 74866.673(16) | 74867.500(16) | 74868.223(17) |
| $2s2p^2$ $^2S_{1/2}$ | $2p^3$ $^2P°_{3/2}$ | 138.39947(6) | | | −2260(53) | 0.485(11) | 0.899(21) | 72254.605(30) | 72255.085(32) | 72255.504(37) |
| $2s2p^2$ $^2S_{1/2}$ | $2p^3$ $^2P°_{1/2}$ | 138.43540(6) | | | −2256(50) | 0.484(11) | 0.897(20) | 72235.855(30) | 72236.334(32) | 72236.752(36) |
| $2s^23s$ $^2S_{1/2}$ | $2s^27p$ $^2P°_{3/2}$ | 142.4325(4) | | | −798(46) | 0.171(10) | 0.318(18) | 70208.68(19) | 70208.85(19) | 70209.00(19) |
| $2s^23s$ $^2S_{1/2}$ | $2s^27p$ $^2P°_{1/2}$ | 142.4345(5) | | | −798(46) | 0.171(10) | 0.318(18) | 70207.71(24) | 70207.88(24) | 70208.03(24) |

**Table 5**
(Continued)

| Lower Level | Upper Level | λ[a] (C II)nm | $K_{MS}$ (inp)[b] (GHz u) | Ref.[c] | $K_{MS}$ (opt)[d] (GHz u) | IS(13−12)cm$^{-1}$ | IS(14−12)cm$^{-1}$ | σ($^{12}$C II)cm$^{-1}$ | σ($^{13}$C II)cm$^{-1}$ | σ($^{14}$C II)cm$^{-1}$ |
|---|---|---|---|---|---|---|---|---|---|---|
| $2s2p^2$ $^2D_{5/2}$ | $2p^3$ $^4S°_{3/2}$ | 149.0398(7) | | | −3229(82) | 0.693(18) | 1.28(3) | 67096.19(30) | 67096.88(30) | 67097.48(30) |
| $2s2p^2$ $^2D_{3/2}$ | $2p^3$ $^4S°_{3/2}$ | 149.0454(7) | | | −3231(82) | 0.693(18) | 1.29(3) | 67093.59(30) | 67094.28(30) | 67094.88(30) |
| $2s^23s$ $^2S_{1/2}$ | $2s^26p$ $^2P°_{3/2}$ | 150.4767(3) | | | −829(46) | 0.178(10) | 0.330(18) | 66455.47(15) | 66455.64(15) | 66455.80(15) |
| $2s^23s$ $^2S_{1/2}$ | $2s^26p$ $^2P°_{1/2}$ | 150.4772(4) | | | −829(46) | 0.178(10) | 0.330(18) | 66455.26(18) | 66455.43(18) | 66455.59(18) |
| $2s2p^2$ $^2S_{1/2}$ | $2s^24p$ $^2P°_{3/2}$ | 151.44422(6) | | | 2396(33) | −0.514(7) | −0.953(13) | 66030.915(30) | 66030.407(31) | 66029.962(33) |
| $2s2p^2$ $^2S_{1/2}$ | $2s^24p$ $^2P°_{1/2}$ | 151.45950(7) | | | 2395(33) | −0.514(7) | −0.953(13) | 66024.255(30) | 66023.747(31) | 66023.303(33) |
| $2s^23s$ $^2S_{1/2}$ | $2s2p(^3P°)3s$ $^2P_{3/2}$ | 163.24966(11) | | | −3291(99) | 0.706(21) | 1.31(4) | 61255.863(40) | 61256.561(45) | 61257.172(56) |
| $2s^23s$ $^2S_{1/2}$ | $2s2p(^3P°)3s$ $^2P_{1/2}$ | 163.30023(14) | | | −3290(99) | 0.706(21) | 1.31(4) | 61236.893(50) | 61237.591(54) | 61238.202(64) |
| $2s^23s$ $^2S_{1/2}$ | $2s^25p$ $^2P°_{3/2}$ | 170.19222(9) | | | −2924(64) | 0.628(14) | 1.16(3) | 58757.083(30) | 58757.704(33) | 58758.247(39) |
| $2s^23s$ $^2S_{1/2}$ | $2s^25p$ $^2P°_{1/2}$ | 170.21360(10) | | | −2924(64) | 0.628(14) | 1.16(3) | 58749.703(40) | 58750.324(42) | 58750.867(48) |
| $2s2p^2$ $^2P_{1/2}$ | $2p^3$ $^2P°_{3/2}$ | 172.04573(14) | | | −1864(59) | 0.400(13) | 0.742(23) | 58124.076(50) | 58124.472(52) | 58124.817(55) |
| $2s2p^2$ $^2P_{1/2}$ | $2p^3$ $^2P°_{1/2}$ | 172.10127(14) | | | −1860(56) | 0.399(12) | 0.740(22) | 58105.326(50) | 58105.721(51) | 58106.066(55) |
| $2s2p^2$ $^2P_{3/2}$ | $2p^3$ $^2P°_{3/2}$ | 172.16839(13) | | | −1862(59) | 0.400(13) | 0.741(23) | 58082.666(40) | 58083.061(42) | 58083.406(46) |
| $2s2p^2$ $^2P_{3/2}$ | $2p^3$ $^2P°_{1/2}$ | 172.22401(13) | | | −1858(56) | 0.399(12) | 0.739(22) | 58063.916(50) | 58064.310(51) | 58064.655(55) |
| $2s2p^2$ $^2D_{5/2}$ | $2s^23p$ $^2P°_{3/2}$ | 176.03944(5) | | | 3836(20) | −0.823(4) | −1.526(8) | 56805.460(15) | 56804.645(16) | 56803.934(17) |
| $2s2p^2$ $^2D_{3/2}$ | $2s^23p$ $^2P°_{3/2}$ | 176.04727(5) | | | 3833(20) | −0.823(4) | −1.525(8) | 56802.935(17) | 56802.121(18) | 56801.410(19) |
| $2s2p^2$ $^2D_{3/2}$ | $2s^23p$ $^2P°_{1/2}$ | 176.08182(5) | | | 3864(21) | −0.829(4) | −1.537(8) | 56791.789(17) | 56790.968(18) | 56790.251(19) |
| $2s^23s$ $^2S_{1/2}$ | $2p^3$ $^2P°_{3/2}$ | 191.53194(11) | | | −5473(60) | 1.174(13) | 2.177(24) | 52210.598(30) | 52211.760(33) | 52212.775(38) |
| $2s^23s$ $^2S_{1/2}$ | $2p^3$ $^2P°_{1/2}$ | 191.60077(11) | | | −5469(58) | 1.174(12) | 2.176(23) | 52191.848(30) | 52193.009(32) | 52194.023(38) |
| $2s2p^2$ $^2P_{1/2}$ | $2s^24p$ $^2P°_{3/2}$ | 192.67682(16) | | | 2792(41) | −0.599(9) | −1.111(16) | 51900.386(40) | 51899.794(41) | 51899.276(43) |
| $2s2p^2$ $^2P_{1/2}$ | $2s^24p$ $^2P°_{1/2}$ | 192.70155(17) | | | 2791(41) | −0.599(9) | −1.110(16) | 51893.726(50) | 51893.134(51) | 51892.616(53) |
| $2s2p^2$ $^2P_{3/2}$ | $2s^24p$ $^2P°_{3/2}$ | 192.83067(16) | | | 2794(41) | −0.600(9) | −1.112(16) | 51858.976(40) | 51858.383(41) | 51857.865(43) |
| $2s2p^2$ $^2P_{3/2}$ | $2s^24p$ $^2P°_{1/2}$ | 192.85545(17) | | | 2793(41) | −0.599(9) | −1.111(16) | 51852.316(50) | 51851.723(51) | 51851.205(53) |
| $2s^23p$ $^2P°_{1/2}$ | $2s^26s$ $^2S_{1/2}$ | 201.7926(3) | | | −332(37) | 0.071(8) | 0.132(15) | 49539.849(80) | 49539.920(80) | 49539.981(81) |
| $2s^23p$ $^2P°_{3/2}$ | $2s^26s$ $^2S_{1/2}$ | 201.8380(3) | | | −301(37) | 0.065(8) | 0.120(15) | 49528.699(80) | 49528.763(80) | 49528.819(81) |
| $2s^23s$ $^2S_{1/2}$ | $2s^24p$ $^2P°_{3/2}$ | 217.38503(12) | | | −817(43) | 0.175(9) | 0.325(17) | 45986.910(24) | 45987.084(26) | 45987.235(29) |
| $2s^23s$ $^2S_{1/2}$ | $2s^24p$ $^2P°_{1/2}$ | 217.41653(14) | | | −818(43) | 0.176(9) | 0.325(17) | 45980.248(30) | 45980.422(31) | 45980.574(35) |
| $2s2p^2$ $^2S_{1/2}$ | $2p^3$ $^4S°_{3/2}$ | 219.5543(15) | | | −4205(81) | 0.902(17) | 1.67(3) | 45532.59(30) | 45533.48(30) | 45534.26(30) |
| $2s^22p$ $^2P°_{1/2}$ | $2s2p^2$ $^4P_{5/2}$ | 232.1995(16) | −3669(15) | B06,K07,J10 | −3669(15) | 0.787(3) | 1.460(6) | 43053.19(30) | 43053.97(30) | 43054.65(30) |
| $2s^22p$ $^2P°_{1/2}$ | $2s2p^2$ $^4P_{3/2}$ | 232.3522(16) | −3669(15) | B07,K07,J10 | −3669(15) | 0.787(3) | 1.460(6) | 43024.89(30) | 43025.67(30) | 43026.35(30) |
| $2s^22p$ $^2P°_{1/2}$ | $2s2p^2$ $^4P_{1/2}$ | 232.4716(15) | −3670(15) | B07,K07,J10 | −3670(15) | 0.788(3) | 1.460(6) | 43002.79(30) | 43003.57(30) | 43004.25(30) |
| $2s^22p$ $^2P°_{3/2}$ | $2s2p^2$ $^4P_{5/2}$ | 232.5419(17) | | | −3668(15) | 0.787(3) | 1.459(6) | 42989.79(30) | 42990.57(30) | 42991.25(30) |
| $2s^22p$ $^2P°_{3/2}$ | $2s2p^2$ $^4P_{3/2}$ | 232.6951(16) | | | −3668(15) | 0.787(3) | 1.459(6) | 42961.49(30) | 42962.27(30) | 42962.95(30) |
| $2s^22p$ $^2P°_{3/2}$ | $2s2p^2$ $^4P_{1/2}$ | 232.8149(15) | | | −3669(15) | 0.787(3) | 1.460(6) | 42939.39(30) | 42940.17(30) | 42940.85(30) |
| $2s^23p$ $^2P°_{1/2}$ | $2s^25s$ $^2S_{1/2}$ | 240.1755(6) | | | −193(37) | 0.041(8) | 0.077(15) | 41623.56(10) | 41623.60(10) | 41623.64(10) |
| $2s^23p$ $^2P°_{3/2}$ | $2s^25s$ $^2S_{1/2}$ | 240.2398(6) | | | −162(37) | 0.035(8) | 0.064(15) | 41612.41(10) | 41612.44(10) | 41612.47(10) |
| $2s2p^2$ $^2P_{1/2}$ | $2p^3$ $^2D°_{3/2}$ | 250.9126(4) | −4522(93) | B57 | −4574(72) | 0.982(15) | 1.82(3) | 39842.510(60) | 39843.481(62) | 39844.329(66) |
| $2s2p^2$ $^2P_{3/2}$ | $2p^3$ $^2D°_{3/2}$ | 251.1736(3) | | | −4571(78) | 0.981(17) | 1.82(3) | 39801.100(50) | 39802.070(53) | 39802.918(59) |
| $2s2p^2$ $^2P_{3/2}$ | $2p^3$ $^2D°_{5/2}$ | 251.2060(4) | −4456(75) | B57 | −4516(63) | 0.969(14) | 1.797(25) | 39795.980(70) | 39796.939(71) | 39797.776(74) |
| $2s2p^2$ $^2S_{1/2}$ | $2s^23p$ $^2P°_{3/2}$ | 283.66987(12) | 2852.7(93) | B57 | 2859.3(83) | −0.6136(18) | −1.138(3) | 35241.889(14) | 35241.281(14) | 35240.751(14) |
| $2s2p^2$ $^2S_{1/2}$ | $2s^23p$ $^2P°_{1/2}$ | 283.75962(12) | 2902(14) | B57 | 2890(12) | −0.6202(25) | −1.150(5) | 35230.743(15) | 35230.129(15) | 35229.593(16) |
| $2s2p^2$ $^2P_{1/2}$ | $2p^3$ $^4S°_{3/2}$ | 318.358(3) | | | −3809(85) | 0.817(18) | 1.52(3) | 31402.09(30) | 31402.90(30) | 31403.61(30) |
| $2s2p^2$ $^2P_{3/2}$ | $2p^3$ $^4S°_{3/2}$ | 318.779(3) | | | −3807(85) | 0.817(18) | 1.51(3) | 31360.59(30) | 31361.40(30) | 31362.11(30) |
| $2s^24s$ $^2S_{1/2}$ | $2s^27p$ $^2P°_{3/2}$ | 338.7447(21) | | | −555(46) | 0.119(10) | 0.221(18) | 29512.28(19) | 29512.40(19) | 29512.50(19) |
| $2s^24s$ $^2S_{1/2}$ | $2s^27p$ $^2P°_{1/2}$ | 338.756(3) | | | −555(46) | 0.119(10) | 0.221(18) | 29511.31(24) | 29511.43(24) | 29511.53(24) |
| $2s^24s$ $^2S_{1/2}$ | $2s^26p$ $^2P°_{3/2}$ | 388.1028(22) | | | −586(46) | 0.126(10) | 0.233(18) | 25759.07(15) | 25759.19(15) | 25759.30(15) |
| $2s^24s$ $^2S_{1/2}$ | $2s^26p$ $^2P°_{1/2}$ | 388.106(3) | | | −586(46) | 0.126(10) | 0.233(18) | 25758.86(18) | 25758.98(18) | 25759.09(18) |

**Table 5**
(Continued)

| Lower Level | Upper Level | $\lambda^a$ (C II)nm | $K_{MS}$ (inp)[b] (GHz u) | Ref.[c] | $K_{MS}$ (opt)[d] (GHz u) | IS(13−12)cm$^{-1}$ | IS(14−12)cm$^{-1}$ | $\sigma(^{12}$C II)cm$^{-1}$ | $\sigma(^{13}$C II)cm$^{-1}$ | $\sigma(^{14}$C II)cm$^{-1}$ |
|---|---|---|---|---|---|---|---|---|---|---|
| $2s^23p\ ^2P°_{1/2}$ | $2s^24s\ ^2S_{1/2}$ | 391.8973(4) | | | 80(32) | −0.017(7) | −0.032(13) | 25509.667(23) | 25509.650(24) | 25509.635(26) |
| $2s^23p\ ^2P°_{3/2}$ | $2s^24s\ ^2S_{1/2}$ | 392.0686(4) | | | 111(32) | −0.024(7) | −0.044(13) | 25498.521(23) | 25498.498(24) | 25498.477(26) |
| $2s^23s\ ^2S_{1/2}$ | $2p^3\ ^4S°_{3/2}$ | 392.221(5) | | | −7418(85) | 1.592(18) | 2.95(3) | 25488.58(30) | 25490.16(30) | 25491.53(30) |
| $2s^23d\ ^2D_{3/2}$ | $2p^3\ ^2P°_{3/2}$ | 430.9317(5) | | | −5525(76) | 1.186(16) | 2.20(3) | 23198.997(30) | 23200.171(34) | 23201.196(43) |
| $2s^23d\ ^2D_{5/2}$ | $2p^3\ ^2P°_{3/2}$ | 430.9586(6) | | | −5525(76) | 1.186(16) | 2.20(3) | 23197.547(30) | 23198.721(34) | 23199.746(43) |
| $2s^23d\ ^2D_{3/2}$ | $2p^3\ ^2P°_{1/2}$ | 431.2804(6) | | | −5521(74) | 1.185(16) | 2.20(3) | 23180.237(30) | 23181.410(34) | 23182.434(42) |
| $2s2p^2\ ^2P_{1/2}$ | $2s^23p\ ^2P°_{3/2}$ | 473.5464(8) | | | 3255(30) | −0.698(6) | −1.295(12) | 21111.357(40) | 21110.666(41) | 21110.062(42) |
| $2s2p^2\ ^2P_{1/2}$ | $2s^23p\ ^2P°_{1/2}$ | 473.7966(8) | | | 3286(30) | −0.705(6) | −1.307(12) | 21100.207(40) | 21099.510(41) | 21098.900(42) |
| $2s2p^2\ ^2P_{3/2}$ | $2s^23p\ ^2P°_{3/2}$ | 474.4771(8) | | | 3257(30) | −0.699(6) | −1.296(12) | 21069.947(40) | 21069.256(41) | 21068.651(42) |
| $2s2p^2\ ^2P_{3/2}$ | $2s^23p\ ^2P°_{1/2}$ | 474.7283(8) | | | 3288(30) | −0.706(6) | −1.308(12) | 21058.797(40) | 21058.099(41) | 21057.489(42) |
| $2s^24s\ ^2S_{1/2}$ | $2s2p(^3P°)3s\ ^2P°_{3/2}$ | 486.2580(8) | | | −3048(99) | 0.654(21) | 1.21(4) | 20559.463(40) | 20560.110(45) | 20560.676(56) |
| $2s^24s\ ^2S_{1/2}$ | $2s2p(^3P°)3s\ ^2P°_{1/2}$ | 486.7071(12) | | | −3048(99) | 0.654(21) | 1.21(4) | 20540.493(50) | 20541.140(54) | 20541.706(64) |
| $2s^24p\ ^2P°_{1/2}$ | $2s^26s\ ^2S_{1/2}$ | 533.2893(21) | | | 163(47) | −0.035(10) | −0.065(19) | 18746.330(80) | 18746.296(81) | 18746.266(82) |
| $2s^24p\ ^2P°_{3/2}$ | $2s^26s\ ^2S_{1/2}$ | 533.4789(22) | | | 162(47) | −0.035(10) | −0.064(19) | 18739.670(80) | 18739.636(81) | 18739.606(82) |
| $2s^24s\ ^2S_{1/2}$ | $2s^25p\ ^2P°_{3/2}$ | 553.5349(8) | | | −2681(64) | 0.575(14) | 1.07(3) | 18060.684(30) | 18061.253(33) | 18061.751(39) |
| $2s^24s\ ^2S_{1/2}$ | $2s^25p\ ^2P°_{1/2}$ | 553.7612(9) | | | −2681(64) | 0.575(14) | 1.07(3) | 18053.304(30) | 18053.873(33) | 18054.371(39) |
| $2s^23d\ ^2D_{3/2}$ | $2s^24p\ ^2P°_{3/2}$ | 588.9277(7) | | | −869(63) | 0.187(14) | 0.35(3) | 16975.307(21) | 16975.492(25) | 16975.653(33) |
| $2s^23d\ ^2D_{5/2}$ | $2s^24p\ ^2P°_{3/2}$ | 588.9780(8) | | | −869(63) | 0.187(14) | 0.35(3) | 16973.858(23) | 16974.043(27) | 16974.204(34) |
| $2s^23d\ ^2D_{3/2}$ | $2s^24p^2P°_{1/2}$ | 589.1589(8) | | | −870(63) | 0.187(14) | 0.35(3) | 16968.645(24) | 16968.830(27) | 16968.991(35) |
| $2s^23s\ ^2S_{1/2}$ | $2s^23p\ ^2P°_{3/2}$ | 657.80482(8) | | | −354(32) | 0.076(7) | 0.141(13) | 15197.8790(18) | 15197.9541(70) | 15198.020(13) |
| $2s^23s\ ^2S_{1/2}$ | $2s^23p\ ^2P°_{1/2}$ | 658.28761(12) | | | −323(32) | 0.069(7) | 0.129(13) | 15186.7333(30) | 15186.8018(75) | 15186.862(13) |
| $2s^23p\ ^2P°_{1/2}$ | $2s^23d\ ^2D_{3/2}$ | 723.1349(11) | | | 375(56) | −0.080(12) | −0.149(22) | 13824.870(21) | 13824.790(24) | 13824.721(31) |
| $2s^23p\ ^2P°_{3/2}$ | $2s^23d\ ^2D_{5/2}$ | 723.6425(13) | | | 406(56) | −0.087(12) | −0.162(22) | 13815.171(30) | 13815.085(32) | 13815.009(37) |
| $2s^23p\ ^2P°_{3/2}$ | $2s^23d\ ^2D_{3/2}$ | 723.7184(11) | | | 406(56) | −0.087(12) | −0.162(22) | 13813.724(21) | 13813.638(24) | 13813.562(31) |
| $2s^25s\ ^2S_{1/2}$ | $2s^27p\ ^2P°_{3/2}$ | 746.153(12) | | | −282(50) | 0.061(11) | 0.112(20) | 13398.39(21) | 13398.45(21) | 13398.50(21) |
| $2s^25s\ ^2S_{1/2}$ | $2s^27p\ ^2P°_{1/2}$ | 746.207(14) | | | −282(50) | 0.061(11) | 0.112(20) | 13397.40(30) | 13397.46(30) | 13397.51(30) |
| $2p^3\ ^2P°_{1/2}$ | $2s^26s\ ^2S_{1/2}$ | 797.564(5) | | | 4814(60) | −1.033(13) | −1.915(24) | 12534.741(80) | 12533.719(81) | 12532.826(84) |
| $2p^3\ ^2P°_{3/2}$ | $2s^26s\ ^2S_{1/2}$ | 798.760(5) | | | 4818(63) | −1.034(13) | −1.917(25) | 12515.981(80) | 12514.958(81) | 12514.064(84) |
| $2s^24s\ ^2S_{1/2}$ | $2p^3\ ^2P°_{3/2}$ | 868.2535(17) | | | −5230(60) | 1.122(13) | 2.081(24) | 11514.199(22) | 11515.310(25) | 11516.280(32) |
| $2s^24s\ ^2S_{1/2}$ | $2p^3\ ^2P°_{1/2}$ | 869.6701(17) | | | −5226(58) | 1.121(12) | 2.079(23) | 11495.444(23) | 11496.554(26) | 11497.523(32) |
| $2s^24p\ ^2P°_{1/2}$ | $2s^25s\ ^2S_{1/2}$ | 923.104(9) | | | 302(47) | −0.065(10) | −0.120(19) | 10830.04(10) | 10829.98(10) | 10829.92(10) |
| $2s^24p\ ^2P°_{3/2}$ | $2s^25s\ ^2S_{1/2}$ | 923.672(9) | | | 301(47) | −0.065(10) | −0.120(19) | 10823.38(10) | 10823.32(10) | 10823.26(10) |
| $2s^25s\ ^2S_{1/2}$ | $2s^26p\ ^2P°_{3/2}$ | 1036.503(19) | | | −313(50) | 0.067(11) | 0.125(20) | 9645.18(18) | 9645.25(18) | 9645.30(18) |
| $2s^25s\ ^2S_{1/2}$ | $2s^26p\ ^2P°_{1/2}$ | 1036.526(22) | | | −313(50) | 0.067(11) | 0.125(20) | 9644.97(21) | 9645.04(21) | 9645.09(21) |
| $2s^25p\ ^2P°_{1/2}$ | $2s^26s\ ^2S_{1/2}$ | 1672.660(23) | | | 2269(67) | −0.487(14) | −0.90(3) | 5976.875(80) | 5976.393(81) | 5975.972(84) |
| $2s^25p\ ^2P°_{3/2}$ | $2s^26s\ ^2S_{1/2}$ | 1674.728(23) | | | 2269(67) | −0.487(14) | −0.90(3) | 5969.495(80) | 5969.013(81) | 5968.592(84) |
| $2s^26s\ ^2S_{1/2}$ | $2s^27p\ ^2P°_{3/2}$ | 1823.62(7) | | | −143(50) | 0.031(11) | 0.057(20) | 5482.10(20) | 5482.13(20) | 5482.16(20) |
| $2s^26s\ ^2S_{1/2}$ | $2s^27p\ ^2P°_{1/2}$ | 1823.94(8) | | | −143(50) | 0.031(11) | 0.057(20) | 5481.13(25) | 5481.16(25) | 5481.19(25) |
| $2s^24s\ ^2S_{1/2}$ | $2s^24p\ ^2P°_{3/2}$ | 1889.661(9) | | | −574(43) | 0.123(9) | 0.228(17) | 5290.509(30) | 5290.631(31) | 5290.737(35) |
| $2s^24s\ ^2S_{1/2}$ | $2s^24p\ ^2P°_{1/2}$ | 1892.043(11) | | | −575(43) | 0.123(9) | 0.229(17) | 5283.849(30) | 5283.971(31) | 5284.077(35) |
| $2p^3\ ^2P°_{1/2}$ | $2s^25s\ ^2S_{1/2}$ | 2165.24(5) | | | 4953(60) | −1.063(13) | −1.971(24) | 4618.44(10) | 4617.39(10) | 4616.47(10) |
| $2p^3\ ^2P°_{3/2}$ | $2s^25s\ ^2S_{1/2}$ | 2174.06(5) | | | 4957(63) | −1.064(13) | −1.972(25) | 4599.69(10) | 4598.64(10) | 4597.72(10) |
| $2s^25s\ ^2S_{1/2}$ | $2s2p(^3P°)3s\ ^2P°_{3/2}$ | 2249.43(5) | | | −2780(100) | 0.596(21) | 1.10(4) | 4445.57(11) | 4446.16(11) | 4446.68(12) |
| $2s^25s\ ^2S_{1/2}$ | $2s2p(^3P°)3s\ ^2P°_{1/2}$ | 2259.07(6) | | | −2770(100) | 0.595(21) | 1.10(4) | 4426.60(11) | 4427.19(11) | 4427.71(12) |
| $2s2p(^3P°)3s\ ^2P°_{1/2}$ | $2s^26s\ ^2S_{1/2}$ | 2865.59(7) | | | 2640(100) | −0.566(21) | −1.05(4) | 3489.686(90) | 3489.126(92) | 3488.637(98) |
| $2s2p(^3P°)3s\ ^2P°_{3/2}$ | $2s^26s\ ^2S_{1/2}$ | 2881.25(7) | | | 2640(100) | −0.566(21) | −1.05(4) | 3470.716(80) | 3470.156(83) | 3469.667(89) |
| $2s^25s\ ^2S_{1/2}$ | $2s^25p\ ^2P°_{3/2}$ | 5136.6(3) | | | −2408(67) | 0.517(14) | 0.96(3) | 1946.79(10) | 1947.31(10) | 1947.75(10) |

**Table 5**
(Continued)

| Lower Level | Upper Level | $\lambda^a$ (C II)nm | $K_{MS}$ (inp)[b] (GHz u) | Ref.[c] | $K_{MS}$ (opt)[d] (GHz u) | IS(13−12)cm$^{-1}$ | IS(14−12)cm$^{-1}$ | $\sigma(^{12}$C II)cm$^{-1}$ | $\sigma(^{13}$C II)cm$^{-1}$ | $\sigma(^{14}$C II)cm$^{-1}$ |
|---|---|---|---|---|---|---|---|---|---|---|
| $2s^25s\ ^2S_{1/2}$ | $2s^25p\ ^2P°_{1/2}$ | 5156.2(3) | | | −2408(67) | 0.517(14) | 0.96(3) | 1939.41(11) | 1939.93(11) | 1940.37(11) |
| $2s^26s\ ^2S_{1/2}$ | $2s^26p\ ^2P°_{3/2}$ | 5784.1(6) | | | −174(50) | 0.037(11) | 0.069(20) | 1728.89(16) | 1728.93(16) | 1728.96(16) |
| $2s^26s\ ^2S_{1/2}$ | $2s^26p\ ^2P°_{1/2}$ | 5784.8(7) | | | −174(50) | 0.037(11) | 0.069(20) | 1728.68(20) | 1728.72(20) | 1728.75(20) |
| $2s^22p\ ^2P°_{1/2}$ | $2s^22p\ ^2P°_{3/2}$ | 157740.92(5) | −1.38(12) | C86 | −1.38(12) | 0.00030(3) | 0.00055(5) | 63.395087(20) | 63.395380(32) | 63.395637(52) |

**Notes.**
[a] Transition wavelength (Ritz value) for natural carbon is given in standard air between 200 and 2000 nm and in vacuum otherwise.
[b] Total mass shift factor values used as input for the least-squares optimization procedure (see text).
[c] Reference code: B57–Burnett (1957) B06–Berengut et al. (2006) B11–Bubin & Adamowicz (2011) B21–Bubin & Adamowicz (2021) C86–Cooksy et al. (1986) H94–Haridass & Huber (1994); J96–Jönsson et al. (1996); J10–Jönsson et al. (2010) K07–Korol & Kozlov (2007).
[d] Total mass shift factor values derived in this work from the input values using the least-squares optimization procedure (see text).

determined in Table 1 for natural carbon:

$$\sigma_{12} = \sigma_{\text{nat}} - c_{13}\text{IS}(^{13}\text{C} - {}^{12}\text{C}), \quad (2)$$

$$\sigma_{13} = \sigma_{\text{nat}} + c_{12}\text{IS}(^{13}\text{C} - {}^{12}\text{C}), \quad (3)$$

where $c_{13} = 0.0106(10)$ is the average relative abundance of $^{13}$C in natural carbon samples (Meija et al. 2016).

As noted in our article on neutral carbon (Haris & Kramida 2017), the uncertainty of the IS($^{13}$C–$^{12}$C) values computed by Berengut et al. (2006) is estimated to be 0.004 cm$^{-1}$ (M. G. Kozlov 2016, private communication). For neutral carbon, this estimate was recently validated by the high-precision measurements of Lai et al. (2020). We assume that this estimate is also valid for C II. It corresponds to a 19 GHz·u uncertainty in the mass-shift factor $K_{\text{MS}}$ defined by the following equation:

$$\delta\nu^{A',A} \equiv \nu^{A'} - \nu^A = K_{\text{MS}}\left(\frac{1}{M_{A'}} - \frac{1}{M_A}\right), \quad (4)$$

where $M_{A'}$ and $M_A$ are the nuclear masses (in atomic mass units, u) of the isotopes with atomic mass numbers $A'$ and $A$; $\delta\nu^{A',A}$ is the difference of transition frequencies between the two isotopes. Different versions of Equation (4) exist in the literature. Some of them replace the nuclear masses with atomic mass numbers (e.g., Berengut et al. 2006), while others add an electron mass to the nuclear masses in the denominators (e.g., Jönsson et al. 2010). At the present level of the precision of experiments and calculations on C II, these differences are negligibly small, and we ignore them here.

Furthermore, the present Equation (4) ignores the so-called field-shift contribution to the IS (see Equation (1) of Korol & Kozlov 2007). This contribution is known to be negligibly small in C I–III at the current level of the precision of experiments and theory.

Most theoretical studies of IS divide the mass shift (MS) into two parts, normal mass shift (NMS) and specific mass shift (SMS), so that $K_{\text{MS}} = K_{\text{NMS}} + K_{\text{SMS}}$. The NMS corresponds to the single-electron term in the Hamiltonian describing the IS, while the SMS corresponds to the two-electron part of the Hamiltonian. In nonrelativistic (NR) theory, the NMS is very simple:

$$K_{\text{NMS}}^{(\text{nr})} = -\nu_\infty m_e, \quad (5)$$

where $m_e$ is the electron's mass and $\nu_\infty$ is the transition frequency computed with an infinite mass of the nucleus. Thus, only the SMS part is challenging to compute. Korol & Kozlov (2007) convincingly showed that, in a consistent relativistic quantum electrodynamics theory, the single-electron part of the Hamiltonian loses its simplicity, and division of the MS into NMS and SMS offers no advantage. Nevertheless, the numerical results of Korol & Kozlov (2007) demonstrate that the difference between relativistic and NR forms is on the order of a few GHz·u for both $K_{\text{SMS}}$ and $K_{\text{NMS}}$, which is well below the precision of their theory for C II. Their Table 3 indicates that the accuracy of their calculation for C II is slightly worse than that of Berengut et al. (2006), at least for the $2s2p^2$ $^2S_{1/2}$–$2s^23p$ $^2P°_{1/2,3/2}$ transitions. According to an evaluation made by Korol & Kozlov for their data on C III, C IV, and a few other spectra, uncertainties of their results vary greatly depending on the transition. The only means of estimating the uncertainties available to us is a statistical analysis based on comparisons with other data. This type of analysis allows for an estimation of only an average uncertainty for all transitions involved. For the C II $K_{\text{MS}}$ values of Korol & Kozlov (2007), our estimated average uncertainty is 25 GHz·u.

As mentioned above, the most precise theoretical data on the C II $K_{\text{MS}}$ factors are provided by Berengut et al. (2006). Their precision is verified to be about 19 GHz·u. Together with the scarce experimental IS data from Burnett (1957) and Cooksy et al. (1986), they served as the base for the evaluation of precision of other calculations.

From old NR calculations of Jönsson et al. (1996), only a few $K_{\text{MS}}$ values are available for C II. Our estimate of their average uncertainty is 40 GHz·u. For the later calculations of Bubin & Adamowicz (2011), the limited amount of data allowed only a rough estimate of the average uncertainty about 50 GHz·u. We remind the reader that all of our $K_{\text{MS}}$ estimations are made for transitions from the ground state. More recently, Hornyák et al. (2020) have published the results of their improved calculation for the five lowest even-parity $^2S$ levels. To utilize the improved accuracy of the latter work, a similarly accurate calculation of the $2s^22p$ $^2P°$ ground term is needed. Such calculation has recently been done by S. Bubin & L. Adamowicz (2021, private communication). These authors have kindly provided their preliminary results to us prior to publication. These new data reproduce the results of Hornyák et al. (2020) for the five lowest even-parity $^2S$ levels, but also include the total energies and IS for the fine-structure levels of the eight lowest $^2P°$ terms of odd parity. We estimated the accuracy of all new $K_{\text{MS}}$ results of S. Bubin & L. Adamowicz (2021, private communication) to be about 35 GHz·u on average. We note that, for the three $K_{\text{MS}}$ values in common with the earlier calculation of Bubin & Adamowicz (2011), the new results are in perfect agreement with the old ones.

For relativistic calculations of Jönsson et al. (2010), the average uncertainty of $K_{\text{MS}}$ values turned out to be 55 GHz·u. However, comparison of their results for O IV with those of a similar calculation of Nazé et al. (2014) indicates a strong correlation between the magnitude of $K_{\text{SMS}}$ and the error in $K_{\text{MS}}$ (here, the error was estimated as a discrepancy between the two compared calculations). This trend is confirmed by the comparison of the $K_{\text{MS}}$ values computed by Jönsson et al. (2010) with the more accurate optimized values (derivation of the latter is described further below). This comparison is depicted in Figure 2. The rms value of all differences $\Delta K$ shown in the figure is 55 GHz·u. However, their unweighted linear fit shown by the dotted line in Figure 2 indicates that the errors in $K_{\text{SMS}}$ values that are outside of the range of available comparisons may be larger. This was accounted for when we assigned the uncertainties to $K_{\text{MS}}$ values that are available from Jönsson et al. (2010) only.

When all theoretical values of $K_{\text{MS}}$ received estimates of uncertainties, it became possible to derive weighted means for the values having multiple determinations. These weighted means and their uncertainties are specified in Table 5. In total, theoretical $K_{\text{MS}}$ values are available for 35 excited levels of C II.

As explained in the beginning of this section, experimental values of IS were measured for eight transitions (Burnett 1957; Cooksy et al. 1986; Haridass & Huber 1994). Only one of these transitions is from the ground level (Cooksy et al. 1986), and it simply replaces the much less accurate theoretical value of

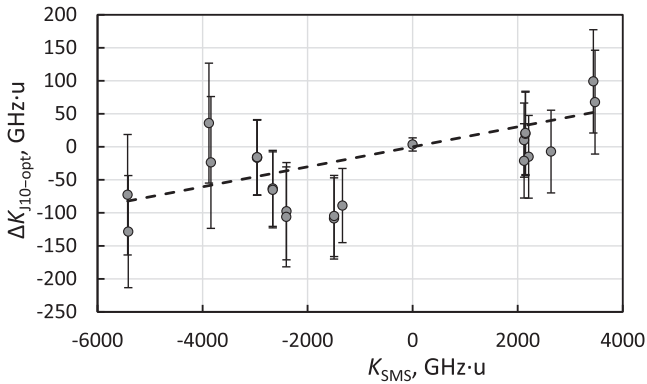

**Figure 2.** Deviations of $K_{MS}$ values computed by Jönsson et al. (2010) from the optimized values plotted against specific mass shift factor $K_{SMS}$. The dotted line is an unweighted linear fit to the data points. The error bars are a sum in quadrature of the uncertainties of the optimized values and estimated uncertainties of Jönsson et al. (2010).

Jönsson et al. (2010). To derive a set of optimized IS values, the experimental data for the remaining seven transitions must be combined with the 35 theoretical values in a self-consistent way. We did it with the help of the least-squares level-optimization code LOPT (Kramida 2011). This is the same code that we used to optimize the energy levels of the natural C II spectrum. The possibility of doing so is explained by the fact that the $K_{MS}$ values obey the same arithmetic rules as energy values: the $K_{MS}$ value for a transition between two excited energy levels equals the difference of $K_{MS}$ values of the upper and lower levels, just the same as transition energy equals the difference between the energies of the upper and lower levels.

To obtain the $K_{MS}$ values for transitions studied by Burnett (1957), we calculated weighted mean averages of the values obtained from his measured IS($^{13}$C–$^{12}$C) and IS($^{14}$C–$^{12}$C) values. These mean averages, as well as all other data used as input for the least-squares optimization procedure, are given in the column "$K_{MS}$ (inp)" of Table 5. The final optimized values produced by the LOPT code are given in the column "$K_{MS}$ (opt)." From those, we calculated the IS values and the wavenumbers for each isotope.

It should be noted that some of the theoretical IS values calculated by different authors are incompatible with each other if the authors' stated uncertainties are applied. For example, S. Bubin & L. Adamowicz (2021, private communication) assess the uncertainties of several of their IS values being so small that the corresponding $K_{MS}$ values would have uncertainties of <5 GHz·u. This includes the $2s2p^2\ ^2S_{1/2}$–$2s^23p\ ^2P°_{1/2,3/2}$ transitions precisely measured by Burnett (1957). Those measurements are in very good agreement with the calculation of Berengut et al. (2006), but are incompatible with the results of S. Bubin & L. Adamowicz (2021, private communication), the differences being $3\sigma$ for the $^2S_{1/2}$–$^2P°_{1/2}$ transition and $9\sigma$ for the $^2S_{1/2}$–$^2P°_{3/2}$ transition, where $\sigma$ is the combined uncertainty of the difference. The stated uncertainty of the calculated $K_{MS}$ values of both transitions is 4 GHz·u, so the $\sigma$ values are dominated by the experimental uncertainties of Burnett (1957; see Table 5). As mentioned above, our much greater adopted values of the uncertainties of the IS values of S. Bubin & L. Adamowicz (2021, private communication) are a consequence of our attempt to reconcile their results with the sparse experimental and other theoretical data. More high-precision calculations using different methods are needed to firmly establish the magnitude of the IS in C II. Also, new high-precision measurements of the IS in the $2s2p^2\ ^2S_{1/2}$–$2s^23p\ ^2P°_{1/2,3/2}$ transitions are needed to resolve the discrepancy with the calculations. In particular, the significant difference between the IS values for the two fine-structure levels of the $2s^23p\ ^2P°$ term observed by Burnett (1957) contradicts all theoretical calculations quoted above, which predict this difference to be smaller than 0.001 cm$^{-1}$ in the IS($^{13}$C–$^{12}$C).

## 5. Hyperfine Structure in $^{13}$C

The only available experimental data on the hyperfine structure (hfs) in C II were obtained by Cooksy et al. (1986) for the magnetic-dipole fine-structure transition within the ground term of $^{13}$C II. As discussed in the previous sections, they determined the center of gravity frequency of this transition in $^{13}$C, 1900545.8(11) MHz, using the laser magnetic resonance method. By using the LSF of their observational data, they also determined a combination of the magnetic-dipole hfs constants $A^{hf}$ of the $^2P°\ J = 1/2$ and $J = 3/2$ ground-term levels, $A^{hf}_{comb} = (A^{hf}_{1/2} - 3A^{hf}_{3/2})/4$. This term equals the offset between the center of gravity of the fine-structure transition and the strongest hfs component, $F'$–$F = 2$–1. Its reported value is 80.3 (4) MHz (note that we give $1\sigma$ uncertainties instead of the original $2\sigma$ values of Cooksy et al. 1986). To predict the frequencies of the remaining two weak hfs components, $F'$–$F = 1$–1 and 1–0, Cooksy et al. (1986) employed restricted Hartree–Fock calculations of Schaefer & Klemm (1970), from which they extracted the values of the $A^{hf}_{1/2}$ and $A^{hf}_{3/2}$ hfs constants assumed to be accurate within 2% or better.

Nowadays, the more accurate calculations of Jönsson et al. (1996), Jönsson et al. (2010) are available for this purpose. Neither of these papers specifies uncertainties for their calculated $A^{hf}$ values, except for some general statements. Jönsson et al. (1996) stated that their results for the hfs constants calculated with the multiconfiguration Hartree–Fock (MCHF) method seem to be converged to within 1% with respect to the increasing active set of orbitals. Jönsson et al. (2010) stated that their results obtained in the large-scale fully relativistic multiconfiguration Dirac–Hartree–Fock (MCDHF) calculations are accurate to within a few percent, and in cases where the hfs constants are large, the accuracy may be even better. However, the results of the MCDHF calculation agree with MCHF to better than 0.6%. This allows us to assign small uncertainties to the MCHF and MCDHF calculated $A^{hf}$ constants for the $2s^22p$ and $2s2p^2$ configurations considered in both of these studies. The numerical values of these uncertainties were obtained by quadratic interpolation of the absolute values of the differences between the two calculations as a function of $A^{hf}$. For configurations not considered by Jönsson et al. (1996) ($2p^3$, $2s^23s$, $2s^23p$, $2s^23d$, $2s^24s$, $2s^24p$, and $2s2p3s$), we adopted an arbitrary estimate of 5% as the relative uncertainty of $A^{hf}$. For small values of $A^{hf}$, which were given by Jönsson et al. (2010) with a limited numerical precision, a lower bound of 2 MHz was adopted for the uncertainties.

Table 6 gives the recommended data on hfs constants, frequencies of the hfs transitions, and Landé $g_J$ factors. For energy levels common in the two large-scale calculations of Jönsson et al. (1996), Jönsson et al. (2010), we adopted a weighted average of their $A^{hf}$ values. For the ground term $2s^22p\ ^2P°$, these average values were used as input in a least-squares optimization procedure together with the $A^{hf}_{comb}$ value measured

**Table 6**
Hyperfine Structure of $^{13}$C II and Landé $g_J$ Factors

| Level | $F'$–$F$ | $\Delta E_{F'-F}$[a] (MHz) | $A^{hfs}$[b] (MHz) | $g_J$[c] | Ref.[d] |
|---|---|---|---|---|---|
| $2s^22p\ ^2P^\circ_{1/2}$ | 1 − 0 | 810.1(16) | 810.1(16) | 0.66576(6) | J10,J96,C86 |
| $2s^22p\ ^2P^\circ_{3/2}$ | 2 − 1 | 325.6(13) | 162.8(6) | 1.33412(6) | J10,J96,C86 |
| $2s2p^2\ ^4P_{1/2}$ | 1 − 0 | 2059(7) | 2059(7) | 2.670373 | J10 |
| $2s2p^2\ ^4P_{3/2}$ | 2 − 1 | 1896(5) | 948(3) | 1.734929 | J10 |
| $2s2p^2\ ^4P_{5/2}$ | 3 − 2 | 2799(8) | 933(3) | 1.601283 | J10 |
| $2s2p^2\ ^2D_{5/2}$ | 3 − 2 | 3147(6) | 1049(2) | 1.200373 | J10,J96 |
| $2s2p^2\ ^2D_{3/2}$ | 2 − 1 | −959(3) | −479.4(15) | 0.799471 | J10,J96 |
| $2s2p^2\ ^2S_{1/2}$ | 1 − 0 | 3678(15) | 3678(15) | 2.002183 | J10,J96 |
| $2s2p^2\ ^2P_{1/2}$ | 1 − 0 | 620.3(16) | 620.3(16) | 0.665857 | J10,J96 |
| $2s2p^2\ ^2P_{3/2}$ | 2 − 1 | −593(3) | −296.6(15) | 1.334014 | J10,J96 |
| $2s^23s\ ^2S_{1/2}$ | 1 − 0 | 1280(60) | 1280(60) | 2.002263 | J10 |
| $2s^23p\ ^2P^\circ_{1/2}$ | 1 − 0 | 110(6) | 110(6) | 0.665864 | J10 |
| $2s^23p\ ^2P^\circ_{3/2}$ | 2 − 1 | 112(6) | 56(3) | 1.334083 | J10 |
| $2p^3\ ^4S^\circ_{3/2}$ | 2 − 1 | −518(26) | −259(13) | 2.002212 | J10 |
| $2s^23d\ ^2D_{3/2}$ | 2 − 1 | 13.8(16) | 6.9(8) | 0.799521 | J10 |
| $2s^23d\ ^2D_{5/2}$ | 3 − 2 | 26.7(26) | 8.9(9) | 1.200449 | J10 |
| $2p^3\ ^2D^\circ_{5/2}$ | 3 − 2 | 717(36) | 239(12) | 1.200381 | J10 |
| $2p^3\ ^2D^\circ_{3/2}$ | 2 − 1 | 368(18) | 184(9) | 0.799487 | J10 |
| $2s^24s\ ^2S_{1/2}$ | 1 − 0 | 264(13) | 264(13) | 2.002303 | J10 |
| $2s^24p\ ^2P^\circ_{1/2}$ | 1 − 0 | 17.0(21) | 17(2) | 0.665872 | J10 |
| $2s^24p\ ^2P^\circ_{3/2}$ | 2 − 1 | 226(11) | 113(6) | 1.334088 | J10 |
| $2s2p3s\ ^4P^\circ_{1/2}$ | 1 − 0 | 2590(130) | 2590(130) | 2.670410 | J10 |
| $2s2p3s\ ^4P^\circ_{3/2}$ | 2 − 1 | 2990(150) | 1495(75) | 1.734904 | J10 |
| $2s2p3s\ ^4P^\circ_{5/2}$ | 3 − 2 | 3500(180) | 1166(58) | 1.601290 | J10 |
| $2p^3\ ^2P^\circ_{1/2}$ | 1 − 0 | 421(21) | 421(21) | 0.665816 | J10 |
| $2p^3\ ^2P^\circ_{3/2}$ | 2 − 1 | 992(50) | 496(25) | 1.334022 | J10 |

**Notes.**
[a] Energy difference between the hyperfine states with total angular momenta $F'$ and $F$.
[b] Magnetic dipole hfs constant. For discussion of the values and their uncertainties, see text.
[c] Landé factors. The two values with uncertainties are experimental ones from Cooksy et al. (1986). The rest are theoretical from Jönsson et al. (2010).
[d] Reference code: C86–Cooksy et al. (1986) J96–Jönsson et al. (1996); J10–Jönsson et al. (2010).

by Cooksy et al. (1986; see above). Thus, the given $A^{hf}_{1/2}$ and $A^{hf}_{3/2}$ values of this term are semiempirical results. They agree with those used by Cooksy et al. (1986) within 1.5%, which testifies to the good accuracy of the calculations of Schaefer & Klemm (1970). For the ground term, the $g_J$ factors are the experimental ones reported by Cooksy et al. (1986). The rest of the $g_J$ factors are those calculated by Jönsson et al. (2010). For the ground term, they agree with experimental ones on the level of precision of Cooksy et al. (1986), i.e., to within $10^{-4}$.

It should be noted that the relative intensities predicted for the hfs transitions within the ground term by Cooksy et al. (1986) are incorrect. This was first mentioned by Ossenkopf et al. (2013) who gave the correct calculated ratio of 0.625:0.25:0.125 for the $(F'-F) = (2-1):(1-0):(1-1)$ transitions. As noted by Ossenkopf et al. (2013), the incorrect values given by Cooksy et al. (1986) had led to errors in some astrophysical studies of elemental abundances and isotopic ratios.

## 6. Observed Line Intensities

As in many other spectra of light elements, the relative intensities of C II lines reported by different observers are on different scales and do not account for variations of sensitivity of the different equipment used and for different excitation conditions. This makes it difficult or impossible to compare intensities reported by different authors. There is no universal solution to this problem. However, there exists a simple method producing acceptable results (Kramida 2013a, 2013b, 2013c; Haris et al. 2014). It is based on four assumptions: 1) the plasma of the light source is in LTE, so that level populations are described by the Boltzmann law with a certain excitation temperature $T_{\text{exc}}$; 2) the plasma is optically thin; 3) all recorded spectral lines were emitted from the same spatial region of the plasma and at the same time interval; 4) the dependence of sensitivity of the spectrometer on wavelength is smooth in large spectral intervals. The last statement reflects the common situation: the entire spectral range studied in one work is often broken into several smaller ranges, each using a different setup (often even a different spectrometer). Under these assumptions, $T_{\text{exc}}$ can be derived from a Boltzmann plot, and then the wavelength-dependent spectral response function of the instrument can be found by fitting the dependence of the ratio of the calculated intensities to observed intensities by piecewise smooth functions. After the removal of the spectral response function from the observed intensities, the intensities reported by different observers differ only by a different $T_{\text{exc}}$ value corresponding to their light sources and a common multiplicative factor. These corrected intensities are easy to convert to a global uniform scale corresponding to a certain selected excitation temperature. When multiple data exist for the relative intensity of the same line, we take a logarithmic average (i.e., exponent of the mean of logarithms of the values) of all available reduced intensity values, excluding data abnormally deviating from the average.

The emission line intensities given in Table 1 have been derived from observations using the method described above.

**Table 7**
Least-Squares Fitted and Hartree–Fock Parameters of C II

| Configurations[a] | Parameter[a] | LSF | Unc.[b] | Group[c] | HFR[a] | LSF/HFR[a] |
|---|---|---|---|---|---|---|
| Odd parity | | | | | | |
| $1s^22s^22p$ | $E_{av}$ | 7181.8 | 16 | | 0 | |
| $1s^22s^22p$ | $\zeta_{2p}$ | 43.1 | | | 43.1 | 1.0000 |
| $1s^22p^3$ | $E_{av}$ | 157,880.8 | 46 | | 153,223.1 | 1.0304 |
| $1s^22p^3$ | $F^2(2p,2p)$ | 49,025.2 | 107 | | 61,514.4 | 0.7970 |
| $1s^22p^3$ | $\alpha_{2p}$ | −680.3 | 6 | | 0 | |
| $1s^22p^3$ | $\zeta_{2p}$ | 41.6 | | | 41.6 | 1.0000 |
| $1s^22s^23p$–$1s^22s2p3$ s | $R_d^{\,1}(2s,3p;2p,3s)$ | 12892 | 134 | 7 | 22,023.1 | 0.5854 |
| $1s^22s^23p$–$1s^22s2p3$ s | $R_e^{\,0}(2s,3p;2p,3s)$ | 2993.2 | 31 | 7 | 5113.2 | 0.5854 |
| $1s^22s^24p$–$1s^22s2p3$ s | $R_d^{\,1}(2s,4p;2p,3s)$ | 5603.9 | 58 | 7 | 9573.0 | 0.5854 |
| $1s^22s^24p$–$1s^22s2p3$ s | $R_e^{\,0}(2s,4p;2p,3s)$ | 1680.6 | 18 | 7 | 2870.9 | 0.5854 |

**Notes.**
[a] Configurations involved in the calculations and their Slater parameters with their least-squares-fitted (LSF) value and/or pseudorelativistic Hartree–Fock (HFR) or their ratio.
[b] Uncertainty of each parameter represents their standard deviation. Blank for fixed values.
[c] Parameters in each numbered group were linked together with their ratio fixed at the HFR level.

(This table is available in its entirety in machine-readable form.)

The effective excitation temperature derived from observations of Glad (1953), 3.5 eV, was used as the base of our intensity scale. The intensities originally reported by Glad (1953) and by Edlén (1934) on a logarithmic scale have first been converted to a roughly linear scale by using the logarithmic base of 1.3 in the original intensities. The TPs required for the Boltzmann plots are discussed in the next section. Most of them are from the previous NIST compilations (Wiese et al. 1996; Wiese & Fuhr 2007), but some are from our own critically evaluated calculations with Cowan's codes.

An exception from the unified intensity scale has been made for the five intercombination lines near 232 nm described in Section 2.4. In most laboratory light sources such as sparks or hollow cathodes, it is difficult or impossible to observe these lines, as they are three orders of magnitude weaker than the other emission lines of C II in plasmas that are close to LTE. These lines were observed only in spectra of the Sun and nebulae, where the extremely low electron densities allow the upper levels of these lines to accumulate a large population. Intensities given in Table 1 for these lines are based on solar values reported by Doschek et al. (1977). Here, they are reduced to the same excitation temperature of 3.5 eV but are increased by an additional factor of 1000 compared to all other emission lines.

Lines reported in absorption (marked with the character "a" in Table 1) have a different intensity scale. It is based on the absorption oscillator strength values experimentally determined by Recanatini et al. (2001) and by Kjeldsen et al. (2001) and assumes that 20% of the total $C^+$ population is in the metastable $2s2p^2\ ^4P$ states. For lines with an unreported experimental oscillator strength, we used the figures with tracings of the observed spectra given in the above papers to estimate the relative intensities.

## 7. Theoretical Calculations of Energy Levels and Transition Probabilities

To interpret the observed spectra, we used Cowan's suite of atomic structure codes (Cowan 1981; Kramida 2018), which implements the Hartree–Fock method with relativistic corrections and superposition of configurations. In the odd-parity set, we included all $1s^2(2s^2+2p^2)n\ell$ ($n \leqslant 10$, $\ell = p, f, h$; $n \leqslant 9$ for $\ell = k$), $1s^22s2pn\ell(n \leqslant 10$ for $\ell = s$; $n \leqslant 9$ for $\ell = d$), $1s^22p^3$, $1s2s^22p(3s + 3d)$, and $1s2s^22p^3$ configurations; in the even parity, we included the $1s^2(2s^2+2p^2)n\ell$ ($n \leqslant 12$ for $\ell = s, d, g$; $n \leqslant 9$ for $\ell = i$), $2s2pn\ell(n \leqslant 9$ for $\ell = p$; $n \leqslant 8$ for $\ell = f$), $1s^22s(2p^2+3s^2+3p^2+3s3d)$, $1s2s^22pnp$ ($n \leqslant 6$), and $1s2p^4$ configurations. A separate calculation was made for the valence-excited Rydberg series, which included $1s^2(2s^2+2p^2)n\ell$ ($n \leqslant 15$) configurations in both parities, but did not include the configurations involving excitation of the 1s electrons. Initial calculations were made with the average energies ($E_{av}$) and spin–orbit parameters ($\zeta_{n\ell}$) taken as 100% of the Hartree–Fock values, while the electrostatic direct and exchange Slater parameters $F^k$ and $G^k$ were scaled to 85% and the configuration interaction parameters $R^k$ to 80% of the Hartree–Fock values. Subsequently, many of these parameters were varied in the LSF procedure that fitted the calculated energy levels to the known experimental values. In the odd parity, 144 experimental levels were fit with 42 free parameters with a standard deviation of 22 cm$^{-1}$. In the even parity, 202 known levels were fit with 43 free parameters with a standard deviation of 152 cm$^{-1}$. All fitted parameter values obtained in the LSF are presented in Table 7.

Our parametric LSF was made under the assumption that Slater parameters describing similar interactions scale similarly in a Rydberg series of configurations. For example, the $F^2(2p,2p)$ parameters describing the electronic structure of the $1s^22p^2$ core of the $1s^22p^2nl$ ($n \geqslant 3$) configurations have been linked in one group (meaning that their ratios within the group remained fixed, and just the scaling coefficient LSF/Hartree–Fock values was allowed to vary). Similarly, the effective parameter $\alpha_{2p}$ of the $2p^2$ subshell was assumed to be the same in all these configurations. The electrostatic exchange parameters $G^1(2p,ns)$ of the same Rydberg series of configurations were also linked in one group (see Table 7). This approach was previously successfully applied to the parametric analyses of many other spectra, e.g., C I (Haris & Kramida 2017), Ne II (Kramida & Nave 2006), Sn II (Haris et al. 2014), Ag II (Kramida 2013b), In II (Kramida 2013c). However, it turned out to be incompatible with some of the identifications of lines observed by Edlén (1934), Recanatini et al. (2001), and Kjeldsen et al. (2001). In Edlén (1934), only one identification has been revised. His assignment of the line at 42.5326 nm to the $1s^22s^22p$ $^2P^\circ$–$1s^22s2p(^1P^\circ)3p$ $^2P$ transition (which he indicated as questionable) has been changed to $1s^22s^22p$ $^2P^\circ$–$1s^22s2p(^3P^\circ)6p$ $^2S$. The latter identification was previously used (Kjeldsen et al. 2001; Recanatini et al. 2001) for the line at 42.627 nm, which we revised to $1s^22s^22p$ $^2P^\circ$–$1s^22s2p(^3P^\circ)6p$ $^2P$. We also revised the identifications of absorption lines at 42.738 and 43.010 nm (Kjeldsen et al. 2001; Recanatini et al. 2001) as specified in Table 1 and identified two of the five unclassified lines listed by Recanatini et al. (2001). We also changed the identifications of four lines listed by Glad (1953) at 282.000, 282.070, 285.800, and 305.685 nm.

With the LSF parameters given in Table 7, TPs (A-values) were recalculated. These new A-values, as well as those from Nahar (2002), Sochi & Storey (2013) and from the recent calculation of Li et al. (2021) were critically evaluated with the help of other previously evaluated results (Wiese et al. 1996; Wiese & Fuhr 2007) that are of either theoretical or experimental origin. The calculations of Kurucz (2014) were also helpful in the assessment of the uncertainties of the new A-values. A detailed description of the evaluation procedure is given elsewhere (Kramida 2013a, 2013b, 2013c; Haris et al. 2014). In total, Table 1 includes 1302 TPs as compared with 628 values previously given by Wiese et al. (1996) and Wiese & Fuhr (2007). Most of the A-values are accurate to better than 25%. For only 175 transitions, the A-values were assigned the poor accuracy categories D+, D, and E. For the description of the accuracy code, see the help pages of the NIST ASD (Kramida et al. 2021) or Table 10 of Haris & Kramida (2017).

### 8. Ionization Energy of C II

As mentioned in the Introduction, Moore (1993) recommended the value 196,664.7 cm$^{-1}$ (with an unspecified uncertainty) for the IE of C II, which was derived by J. Bromander from the data of Glad (1953). Using the computer code POLAR by C. J. Sansonetti (2005, private communication), we were able to reproduce the fit made by Bromander and evaluate its uncertainty as 2.2 cm$^{-1}$.

A more recent value, 196,674(7) cm$^{-1}$, was obtained by Biémont et al. (1999) using an interpolation of the differences between the experimental and theoretical (multiconfiguration Dirac-Fock) data.

The present work significantly extended the experimental data available for highly excited Rydberg levels and thus offers a possibility to determine the IE with improved precision. Among the several observed Rydberg series, the ones with the largest number of accurately determined levels are $1s^22s^2nf$ and $1s^22s^2ng$. In both of these series, the fine-structure splittings were not resolved in the highest members, so only the centers of gravity of their doublet terms can be used for determination of the IE. We used a newly written computer code *fit_Ritz* to simultaneously fit the extended Ritz quantum-defect expansion formulas (Kramida 2013a) for these two series with the common series limit, C III $1s^22s^2$ $^1S_0$. The code determines the uncertainty of the fitted value of the series limit by tracking the changes in the limit value resulting from uncertainties of each input level value.

In the $1s^22s^2nf$ series, the levels with $n \leqslant 7$ are known from the work of Glad (1953). However, the $n = 7$ level had to be excluded from the fit, as it deviated from the fitted value by about 1 cm$^{-1}$ (i.e., by about $5\sigma$). This deviation is probably caused by the Stark effect, which was considerable in Glad's work according to his observations. Transitions from the $nf$ levels with $n = 8$ to 14 were present in the nebula spectrum studied by Sharpee et al. (2003, 2004), except for the $n = 10$ and $n = 13$ levels, transitions from which down to $1s^22s^24d$ were masked by other species. The $1s^22s^24d$–$1s^22s^2nf$ ($n = 8$, 9) transitions were identified by Sharpee et al. (2003). We identified the previously unclassified lines at 401.5931 and 392.4007 nm in their line list as the $n = 11$ and 12 members of this series. The line at 380.2720 nm was assigned by Sharpee et al. (2003) to a transition in S II, but the observed wavelength deviated strongly from the expected (Ritz) value, so this assignment was marked as questionable. Our calculations indicate that most of the intensity of this line is due to the $1s^22s^24d$–$1s^22s^214f$ transition in C II, although some contribution from S II cannot be ruled out. For the fit of this series, we used the four-term extended Ritz formula (see Kramida 2013a).

In the $1s^22s^2ng$ series, the situation is similar. The lower members with $n \leqslant 8$ have been observed by Glad (1953), but only the $n = 5$ level is determined here from his measurements. The higher members of the series appear to be significantly affected by Stark shifts in Glad's experiment. The rest of Glad's measurements for this series have been replaced here by the results of Sharpee et al. (2003), as their wavelengths (reduced to the terrestrial rest frame) are measured more accurately and are not affected by Stark shifts. Sharpee et al. (2003) identified only six $1s^22s^24f$–$1s^22s^2ng$ lines with $n = 6$ to 10 and 14. We found in their line list the $n = 11$ and 12 members of this series and additionally identified five lines of the $1s^22s^25f$–$1s^22s^2ng$ series with $n = 8$ to 12. The three $n = 8$, 9, 10 lines were previously unclassified; for the higher members, we revised the identifications of Sharpee et al. (2003). For the fit of this series, we used the two-term extended Ritz formula (see Kramida 2013a).

As a result of the joint fit of the quantum-defect expansion formulas for the $1s^22s^2nf$ and $1s^22s^2ng$ series, we obtained a value of 196,663.27(11) cm$^{-1}$ for the first ionization limit.

In addition to the $nf$ and $ng$ series, we also identified two lines of the $1s^22s^25g$–$1s^22s^2nh$ series with $n = 8$ and 10. These lines were listed by Sharpee et al. (2003) at 931.8089 nm and 757.4071 nm, respectively (both were unclassified). The $n = 11$ line of this series is predicted to occur at 716.058(11) nm. In the line list of Sharpee et al. (2003), there is a line at 716.0559 nm, but it is identified as the $1s3s$ $^3S$–$1s10p$ $^3P^\circ$ transition in He I. Its observed intensity is much greater than that predicted for the C II $1s^22s^25g$–$1s^22s^211h$ transition, so we conclude that the latter is masked by the He I line in the spectrum of the nebula IC 418. The lower $nh$ levels with $n = 6$ and 7 had been determined by Glad (1953) from the parity-forbidden $1s^22s^24f$–$1s^22s^2nh$ transitions. The very fact of their observation testifies to the presence in Glad's light source of strong electric fields enabling these transitions. Thus, these two levels are likely to have been strongly affected by Stark shifts. This leaves us with only two $nh$ levels that are accurately measured in a field-free (nebula) environment. Although it is insufficient for a fit of the quantum defects along the series, these levels can be used in a fit of the polarization formula (Sansonetti et al. 1981; C. J. Sansonetti 2005, private communication; Kramida 2013a). We fitted these two levels together with the $1s^22s^2ng$ ($n = 5$–12, 14) levels using the code POLAR by C. J. Sansonetti (2005, private communication),

Table 8
Determinations of Ionization Energy of C II

| No. | Series ($2s^2n\ell$) | IE[a] (cm$^{-1}$) | Method[b] | References[c] | Notes[d] |
|---|---|---|---|---|---|
| 1 | $ng + n'$h ($n = 5$–8, $n' = 6, 7$) | 196,664.7(22) | Polar | M93 | G53 |
| 2 | $2p\ ^2P°$ | 196,674(7) | SE (MCDF) | B99 | |
| 3 | $2p\ ^2P°$ | 196,653.7(6) | TH (ECG) | B10, B11 | NR |
| 4 | $ng + n'$f ($n = 5$–12,14; $n' = 4$–6, 8, 9, 11, 12, 14) | 196,663.27(11) | Ritz q.d. | TW | S03 |
| 5 | $ng + n'$h ($n = 5$–12, 14; $n' = 8, 10$) | 196,663.45(20) | Polar | TW | S03 |
| 6 | Recommended value | 196,663.31(10) | Weighted average of nos. 4 and 5. | | |

**Notes.**
[a] The uncertainties given in parentheses after the IE values are either quoted from the original sources or determined in the present work.
[b] Method of determination of the IE: Polar–fit of the polarization formula; Ritz q.d.–fit of Ritz quantum-defect expansion formulas; SE–semiempirical value; TH–ab initio theory; MCDF–multiconfiguration Dirac-Fock; ECG–explicitly correlated Gaussian basis functions.
[c] M93–Moore (1993) B99–Biémont et al. (1999) B10–Bubin et al. (2010); B11–Bubin & Adamowicz (2011) TW–this work.
[d] G53–semiempirical value derived by J. Bromander from the data of Glad (1953), as referred to in Moore (1993), which are affected by Stark shifts; NR–nonrelativistic variational calculations; S03–derived in this work from the data of Sharpee et al. (2003), Sharpee et al. (2004), observed in the IC 418 nebula, which is free of Stark shifts.

which we modified to enable tracking propagation of the uncertainties from the input level values to the final fitted value of the limit. The resulting value of the limit was 196,663.45(20) cm$^{-1}$. In addition to the limit value, the POLAR code also allows determination of interpolated values for the series members excluded from the fit. The interpolated values of the $1s^22s^26h$ and $1s^22s^27h$ levels are 184,464.11(14) cm$^{-1}$ and 187,700.90(12) cm$^{-1}$, respectively. The first one is 2.37(17) cm$^{-1}$ lower and the second 0.66(19) cm$^{-1}$ higher than the experimental values found from the measurements of Glad (1953). This supports our conclusion that Glad's values of these levels are strongly affected by the Stark effect.

The final recommended IE value, 196,663.31(10) cm$^{-1}$, is a weighted mean of our two determinations discussed above. It is equivalent to 24.383143(13) eV.[5] The different determinations discussed above are summarized in Table 8.

Our recommended experimental value agrees well with previous experimental and semiempirical determinations but is more precise by more than an order of magnitude.

The most accurate purely theoretical determination of the IE was made by Bubin et al. (2010), Bubin & Adamowicz (2011). In these two works, they gave NR variational total energies of the ground terms of C III and C II, respectively. The value given in Table 8 referring to these two works corresponds to the difference between these total NR energies, and its uncertainty reflects only the uncertainty of the convergence in their variational calculations. It can be seen that the NR value of the IE differs from the experimental one by about 10 cm$^{-1}$. The relativistic and QED effects were calculated for the ground term of C III (Bubin et al. 2010) but not for C II; they are currently working on deriving these corrections for C II (S. Bubin & L. Adamowicz 2021, private communication) and improving the accuracy of the NR values.

With the IE fixed at the newly determined value, we used the extended Ritz quantum-defect formula to interpolate and extrapolate the values of the $[1s^22s^2]nf$ and $ng$ levels and the polarization formula to interpolate and extrapolate the $nh$, $ni$, and $nk$ levels for $n \leqslant 15$. These interpolated values are given in Table 2 for the unobserved levels of these series. The uncertainties of the extrapolated and interpolated values were found as a combination in quadrature of the uncertainty of the ionization limit, 0.10 cm$^{-1}$, and the uncertainty of the interpolation procedure determined by the fit_Ritz and POLAR codes by propagating experimental uncertainties of the known levels.

In the $[1s^2]2s^2ns$ series, the first member with $n = 3$ is strongly mixed with $2s2p^2\ ^2S$. Highly excited members of this series with $n \geqslant 10$ strongly interact with the $2s2p(^3P°)3p\ ^2S$ level. The exact location of this level is experimentally unknown. Its predicted position varies strongly (between 192,000 and 194,000 cm$^{-1}$) in different theoretical models, but all models agree in that it strongly mixes with the high-$n$ $2s^2ns$ levels, which makes the extrapolations of quantum defects unreliable. Thus, we retained the extrapolated values only for the $n = 8$ and 9 members of this series.

## 9. Comparisons with Previous Reference Data Sets

Before this work, the most widely used sources of reference wavelengths of C II were the compilation of Kelly (1987) and the list of recommended Ritz wavelengths of Kaufman & Edlén (1974). Kelly (1987) lists 313 lines of C II. Of these lines, three represent Ritz wavelengths of the forbidden $2s^22p\ ^2P°$–$2s^23p\ ^2P°$ multiplet, which can be observed only under special conditions. The remaining 310 lines refer to six sources: Edlén (1934), Glad (1953), Herzberg (1958, 1962), Moore (1970), and Junkes et al. (1965). However, only 138 of these 310 wavelengths correspond to those reported in the given references. Moreover, the wavelengths quoted from Herzberg (1962) were in fact not observed but calculated from the Ritz principle (i.e., from the differences of the upper and lower energy levels of each transition). It appears that most of the C II wavelengths given in Kelly (1987) are Ritz values calculated from energy levels determined by Glad (1953). Note that, following Junkes et al. (1965), the latter reference was quoted in Kelly (1987) with a wrong year, 1954, which has caused much confusion in subsequent literature. Of the two lines quoted from Junkes et al. (1965), only one appears in that book, and it is listed as a weak line without an assignment to C II. Most of the carbon lines observed by Junkes et al. (1965) belong to C I, so it is unlikely that this weak line was due to C II. In the present compilation, we ignored this observation, although the observed wavelength, 192.830 nm, agrees well with our Ritz wavelength, 192.83067(16) nm, for the C II transition assigned by Kelly (1987), $2s2p^2\ ^2P_{3/2}$–$2s^24p\ ^2P°_{3/2}$. On average, the wavelengths listed by Kelly (1987) are in fair agreement with our Ritz values given in Table 1. For example, for the wavelengths given by Kelly (1987) with a numerical precision of 0.001 Å (0.0001 nm), the rms difference from our Ritz values is 0.012 Å (0.0012 nm). This is to be expected, since we revised only a few of the identifications of Glad (1953), and most of our optimized energy levels listed in Table 2 agree with his values fairly well. However, nine out of 164 wavelengths given by Kelly with this precision differ substantially from our more accurate Ritz values with the differences ranging from −0.056 to 0.078 Å (−0.0056 to 0.0078 nm). This supports our statement made in Section 2.3: the use of the Kelly (1987) compilation as a source of reference wavelengths is strongly discouraged. It is a great resource for finding the relevant literature, but the wavelengths listed by Kelly should be used with caution.

---
[5] Conversion from the customary units of energy (cm$^{-1}$) used in this work to eV is made with the 2018 CODATA recommended factor, 1 eV = 8065.54393734921 cm$^{-1}$ (see Tiesinga et al. 2020).

The compilation of reference wavelengths by Kaufman & Edlén (1974) is in general a much better source of secondary wavelength standards, especially in the VUV, XUV, and soft X-ray regions. It lists Ritz wavelengths with specified uncertainties, which are in most cases quite reasonable. However, it does contain some errors, as some of the energy levels used to derive these Ritz wavelengths have been revised after that publication. See, for example, the work of Kramida et al. (2017) on Cu II. Kaufman & Edlén (1974) have listed 37 lines of C II between 54.9 and 176.1 nm with stated uncertainties between 0.05 and 0.10 pm (0.5–1 mÅ). To derive the energy levels used to determine these Ritz wavelengths, they combined Glad's data with the more precise measurements of Herzberg (1958). The rms difference of the C II wavelengths listed by Kaufman & Edlén (1974) from our Ritz wavelengths given in Table 1 is 0.05 pm (0.5 mÅ) with the largest difference being 0.15 pm (1.5 mÅ). This constitutes a very good agreement. Uncertainties of our Ritz wavelengths for these 37 lines range from 0.010 to 0.11 pm (0.10–1.1 mÅ), the average being 0.04 pm (0.4 mÅ). This is about a factor of two smaller than the average uncertainty of Kaufman & Edlén (1974) for these wavelengths. Our compilation provides many more similarly precise Ritz wavelengths.

## 10. Conclusion

In this work, we have presented a set of extended and critically evaluated atomic data for singly ionized carbon (C II) based on 597 selected observed lines assigned to 972 electronic transitions. From these lines, a total of 414 energy levels have been established along with accurate Ritz wavelengths with well-defined uncertainties. Among these levels, 350 are optimized in a least-squares minimization procedure based on observed spectral lines, while 64 levels are determined by interpolations and extrapolations using either a quantum-defect expansion formula or a polarization formula. Along with the level values, we give the percentage compositions of the eigenvectors determined in our parametric calculations with Cowan's codes. Relative intensities of the observed emission lines reported by different researchers have been reduced to a common linear scale. The list of observed lines has been supplemented with a few hundred possibly observable lines, for which the Ritz wavelengths have been calculated using the optimized energy levels. Wherever possible, critically assessed TPs have been included in the line list. The present data set includes 1303 TP values, 273 of which have been calculated in the present work. This more than doubles the number of TP values compared with the previously recommended data set (Wiese et al. 1996 Wiese & Fuhr 2007). Using the present extended list of precisely determined energy levels, the principal IE of C II has been determined, which reduces the uncertainty by more than a factor of 20 in comparison with the previously recommended value. With this improvement in IE, many highly excited energy levels of the $1s^22s^2n\ell\ ^2L$ ($n \leqslant 15$, $\ell = s, f, g, h, i, k$) Rydberg series have been accurately interpolated or extrapolated using quantum defect and polarization formulas.

By combining experimental and theoretical data on isotope shifts with the optimized energy levels of natural carbon, we have deduced 35 excited energy levels, as well as wavelengths and isotope shifts for 187 transitions in the spectra of three singly ionized carbon isotopes: $^{12}C^+$, $^{13}C^+$, and $^{14}C^+$.

The above achievements should not deceive the reader into believing that the knowledge of the C II spectrum is presently complete and accurate. On the contrary, we would like to stress that the bulk of the data presented here are based on old laboratory measurements plagued by Stark shifts of unknown but large magnitude (Glad 1953) and on nebular spectra that are prone to poorly controlled Doppler shifts (different for different parts of a nebula) and blending with other species (Sharpee et al. 2003; Sharpee et al. 2004). Comparing with the C I spectrum, the precision of the C II energy levels is worse by 2 to 3 orders of magnitude for most levels. The fine structure is unresolved for most levels, so that crudely calculated theoretical splittings had to be utilized to locate the levels. The present theoretical interpretation of multiple observed Rydberg series should be treated as questionable, especially for autoionizing levels. For the latter, Auger decay rates calculated by different atomic codes vary by orders of magnitude and poorly agree with very scarce and imprecise measurements of absorption line profiles. A few available measurements of IS do not allow for a dependable assessment of the accuracy of several theoretical calculations, whose authors give contradicting estimates of their uncertainties, making their results incompatible with each other. This spectrum needs more experimental and theoretical work.

This work was partially funded by the Astrophysics Research and Analysis program of the National Aeronautics and Space Administration of the USA, grant No. 80HQTR19T0051. The support from Rodrigo Ibacache, NIST, with the smooth data digitization process is gratefully acknowledged. Haris was working under Guest Researcher agreement 131227 with NIST. We thank Dr. S. Bubin and Prof. L. Adamowicz for providing preliminary results of their new calculations prior to publication.

### ORCID iDs

A. Kramida ● https://orcid.org/0000-0002-0788-8087
K. Haris ● https://orcid.org/0000-0002-1341-6297

### References

Audi, G., Kondev, F. G., Wang, M., et al. 2017, ChPhC, 41, 030001
Berengut, J. C., Flambaum, V. V., & Kozlov, M. G. 2006, PhRvA, 73, 012504
Biémont, E., Frémat, Y., & Quinet, P. 1999, ADNDT, 71, 117
Bockasten, K. 1955, Ark Fys, 9, 457
Bowen, I. S. 1927, PhRv, 29, 231
Bowen, I. S., & Ingram, S. B. 1926, PhRv, 28, 444

Boyce, J. C., & Rieke, C. A. 1935, PhRv, 47, 653
Braams, B. J., & Chung, H.-K. 2015, JPhCS, 576, 011001
Brault, J. W. 1978, in Proc. JOSO Workshop 106, Future Solar Optical Observations Needs and Constraints, ed. G. Godoli (Florence: Pubblicazioni della Universita degli Studi di Firenze), 33
Bubin, S., & Adamowicz, L. 2011, JChPh, 135, 214104
Bubin, S., Komasa, J., Stanke, M., & Adamowicz, L. 2010, PhRvA, 81, 052504
Burnett, C. R. 1950, PhRv, 80, 494
Burnett, C. R. 1957, BAPS, 2, 200
Burns, K., Adams, K. B., & Longwell, J. 1950, JOSA, 40, 339
Burns, K., & Walters, F. M., Jr. 1929, PAllO, 6, 159
Burns, K., & Walters, F. M., Jr. 1931, PAllO, 8, 39
Burton, W. M., & Ridgeley, A. 1970, SoPh, 14, 3
Cooksy, A. L., Blake, G. A., & Saykally, R. J. 1986, ApJL, 305, L89
Corrégé, G., & Hibbert, A. 2002, JPhB, 35, 1211
Cowan, R. D. 1981, in The Theory of Atomic Structure and Spectra, ed. A. Kramida (Berkeley, CA: Univ. of California Press)
Curdt, W., Brekke, P., Feldman, U., et al. 2001, A&A, 375, 591
Curdt, W., Feldman, U., Laming, J. M., et al. 1997, A&AS, 126, 281
Ding, R., Rudakov, D. L., Stangeby, P. C., et al. 2017, NucFu, 57, 056016
Dobbie, J. C. 1938, The Spectrum of Fe II, Ann. Solar Phys. Obs. Cambridge (Vol 5, Part 1) (Cambridge: Cambridge Univ. Press)
Doschek, G. A., Feldman, U., & Cohen, L. 1977, ApJS, 33, 101
Edlén, B. 1933a, ZPhy, 84, 746
Edlén, B. 1933b, ZPhy, 85, 85
Edlén, B. 1934, Nova Acta Reg. Soc. Sci. Upsaliensis Ser. IV, 9, 1
Edlén, B. 1936, ZPhy, 98, 561
Edlén, B. 1963, RPPh, 26, 181
Fang, X., & Liu, X.-W. 2011, MNRAS, 415, 181
Feldman, U., Behring, W. E., Curdt, W., et al. 1997, ApJS, 113, 195
Fowler, A., & Selwyn, E. W. H. 1928, RSPSA, 120, 312
Froese Fischer, C., & Tachiev, G. 2004, ADNDT, 87, 1
García-Rojas, J., Peña, M., Morisset, C., Mesa-Delgado, A., & Ruiz, M. T. 2012, A&A, 538, A54
Glad, S. 1953, Ark Fys, 7, 7
Gnaciñski, P. 2009, AcA, 59, 325
Graf, A. T., May, M. J., Beiersdorfer, P., & Terry, J. L. 2011, CaJPh, 89, 615, [Erratum: 2011, CaJPh, 89, 825]
Griesmann, U., & Kling, R. 2000, ApJL, 536, L113
Hakalla, R., Niu, M. L., Field, R. W., et al. 2016, RSCAd, 6, 31588
Haridass, C., & Huber, K. P. 1994, ApJ, 420, 433
Haris, K., & Kramida, A. 2017, ApJS, 233, 16
Haris, K., Kramida, A., & Tauheed, A. 2014, PhyS, 89, 115403
Henning, T., & Salama, F. 1998, Sci, 282, 2204
Herzberg, G. 1958, RSPSA, 248, 309
Herzberg, G. 1962, IAUTA, 11, 97
Hill, F., Bogart, R. S., Davey, A., et al. 2004, Proc. SPIE, 5493, 163
Hornyák, I., Adamowicz, L., & Bubin, S. 2020, PhRvA, 102, 062825
Huber, M. C. E., Sandeman, R., & Tozzi, G. P. 1984, PhST, 8, 95
Iijima, T., & Nakanishi, H. 2008, A&A, 482, 865
Isler, R. C., Wood, R. W., Klepper, C. C., et al. 1997, PhPl, 4, 356
Jannitti, E., Gaye, M., Mazzoni, M., Nicolosi, P., & Villoresi, P. 1993, PhRvA, 47, 4033
Jönsson, P., Froese Fischer, C., & Godefroid, M. R. 1996, JPhB, 29, 2393
Jönsson, P., Li, J.-G., Gaigalas, G., & Dong, C.-Z. 2010, ADNDT, 96, 271
Junkes, J., Salpeter, E. W., & Milazzo, G. 1965, Atomic Spectra in the Vacuum Ultraviolet from 2250 to 1100 Å, Part One–Al, C, Cu, Fe, Ge, Hg, Si, ($H_2$) (Vatican City: Specola Vaticana)
Kaler, J. B. 1976, ApJS, 31, 517
Kaufman, V., & Edlén, B. 1974, JPCRD, 3, 825
Kelly, R. L. 1987, Atomic and Ionic Spectrum Lines Below 2000 Angstroms: Hydrogen Through Krypton, J. Phys. Chem. Ref. Data 16, Suppl. 1 (Washington, DC: American Chemical Society) [Erratum: 1998, JPCRD, 17, 953]
Kjeldsen, H., Folkmann, F., Hansen, J. E., et al. 1999, ApJL, 524, L143
Kjeldsen, H., Hansen, J. E., Folkmann, F., et al. 2001, ApJS, 135, 285
Konjević, N., Lesage, A., Fuhr, J. R., & Wiese, W. L. 2002, JPCRD, 31, 819
Korol, V. A., & Kozlov, M. G. 2007, PhRvA, 76, 022103
Kramida, A. 2013a, Fusion Sci. Technol., 63, 313
Kramida, A. 2013b, NISTJ, 118, 168
Kramida, A. 2013c, NISTJ, 118, 52
Kramida, A. 2018, A Suite of Atomic Structure Codes Originally Developed by RD Cowan Adapted for Windows-Based Personal Computers (Gaithersburg, MD: National Institute of Standards and Technology)
Kramida, A., Nave, G., & Reader, J. 2017, Atoms, 5, 9
Kramida, A., Ralchenko, Y., Reader, J. & NIST ASD Team 2021, NIST Atomic Spectra Database, v5.9 (Gaithersburg, MD: National Institute of Standards and Technology), https://physics.nist.gov/asd
Kramida, A. E. 2011, CoPhC, 182, 419
Kramida, A. E., & Nave, G. 2006, EPJD, 39, 331
Kurucz, R. L. 2014, Atomic Line Lists and Energy Level Calculations, http://kurucz.harvard.edu/atoms.html
Lai, K.-F., Ubachs, W., De Oliveira, N., & Salumbides, E. J. 2020, Atoms, 8, 62
Levshakov, S. A., Kozlov, M. G., & Agafonova, I. I. 2020, MNRAS, 498, 3624
Levshakov, S. A., Ng, K.-W., Henkel, C., & Mookerjea, B. 2017, MNRAS, 471, 2143
Li, W., Amarsi, A. M., Papoulia, A., Ekman, J., & Jönsson, P. 2021, MNRAS, 502, 3780
McCarthy, K. J., Zurro, B., Hollmann, E. M., Sánchez, J. H. & the TJ-II team 2016, PhyS, 91, 115601
Meija, J., Coplen, T. B., Berglund, M., et al. 2016, PApCh, 88, 293
Menmuir, S., Giroud, C., Biewer, T. M., et al. 2014, RScI, 85, 11E412
Moore, C. E. 1949, NBS Circular 467, Atomic Energy Levels as Derived from the Analysis of Optical Spectra–Hydrogen through Vanadium (Washington, DC: National Bureau of Standards)
Moore, C. E. 1970, Selected Tables of Atomic Spectra: C I, C II, C III, C IV, C V, C VI (Washington, DC: US Govt. Printing Office)
Moore, C. E. 1993, Tables of Spectra of Hydrogen, Carbon, Nitrogen, and Oxygen Atoms and Ions (Boca Raton, FL: CRC Press)
Müller, A., Borovik, A., Jr., Buhr, T., et al. 2018, PhRvA, 97, 013409
Nahar, S. N. 1995, ApJS, 101, 423, [Erratum: 1996, ApJS, 106, 213]
Nahar, S. N. 2002, ADNDT, 80, 205
Nave, G., & Clear, C. 2021, MNRAS, 502, 5679
Nazé, C., Verdebout, S., Rynkun, P., et al. 2014, ADNDT, 100, 1197
Nicolosi, P., & Villoresi, P. 1998, PhRvA, 58, 4985
Nussbaumer, H., & Storey, P. J. 1981, A&A, 96, 91
Oishi, T., Morita, S., Huang, X. L., Zhang, H. M., & Goto, M. 2014, RScI, 85, 11E415
Ossenkopf, V., Röllig, M., Neufeld, D. A., et al. 2013, A&A, 550, A57
Otsuka, M., Tajitsu, A., Hyung, S., & Izumiura, H. 2010, ApJ, 723, 658
Palmer, B. A., & Engleman, R., Jr. 1983, Atlas of the Thorium Spectrum, Report, LA-9615, (Los Alamos, NM: Los Alamos National Laboratory)
Parenti, S., Vial, J.-C., & Lemaire, P. 2005, A&A, 443, 679
Peck, E. R., & Reeder, K. 1972, JOSA, 62, 958
Penston, M. V., Benvenuti, P., Cassatella, A., et al. 1983, MNRAS, 202, 833
Recanatini, P., Nicolosi, P., & Villoresi, P. 2001, PhRvA, 64, 012509
Risberg, P. 1955, Ark Fys, 9, 483
Rødbro, M., Bruch, R., & Bisgaard, P. 1979, JPhB, 12, 2413
Russell, R. W., Melnick, G., Gull, G. E., & Harwit, M. 1980, ApJL, 240, L99
Saloman, E. B., & Sansonetti, C. J. 2004, JPCRD, 33, 1113
Sansonetti, C. J., Andrew, K. L., & Verges, J. 1981, JOSA, 71, 423
Schaefer, H. F., & Klemm, R. A. 1970, PhRvA, 1, 1063
Scully, S. W. J., Aguilar, A., Emmons, E. D., et al. 2005, JPhB, 38, 1967
Sharpee, B., Baldwin, J. A., & Williams, R. 2004, ApJ, 615, 323
Sharpee, B., Williams, R., Baldwin, J. A., & van Hoof, P. A. M. 2003, ApJS, 149, 157
Sharpee, B. D. 2003, PhD thesis, Univ. of Michigan, An Abundance Study of IC 418 Using High Resolution, Signal-to-Noise Emission Spectra, https://doi.org/10.25335/M5DZ0396M
Smith, S. J., Chutjian, A., & Greenwood, J. B. 1999, PhRvA, 60, 3569
Sochi, T., & Storey, P. J. 2013, ADNDT, 99, 633
Storey, P. J., & Sochi, T. 2013, MNRAS, 430, 599
Tachiev, G., & Froese Fischer, C. 2000, JPhB, 33, 2419
Tiesinga, E., Mohr, P. J., Newell, D. B., & Taylor, B. N. 2020, in The 2018 CODATA Recommended Values of the Fundamental Physical Constants (Web v8.1), Database Developed, Baker, J., Douma, M., & Kotochigova, S. (Gaithersburg, MD, USA: National Institute of Standards and Technology), https://physics.nist.gov/constants
Träbert, E., Gwinner, G., Knystautas, E. J., Tordoir, X., & Wolf, A. 1999, JPhB, 32, L491
van Hoof, P. A. M. 1999, Atomic Line List, v204, http://www.pa.uky.edu/%7Epeter/atomic/
von Hellermann, M. G., Bertschinger, G., Biel, W., et al. 2005, PhST, 120, 19
Whaling, W., Anderson, W. H. C., Carle, M. T., Brault, J. W., & Zarem, H. A. 1995, JQSRT, 53, 1
Wiese, W. L., & Fuhr, J. R. 2007, JPCRD, 36, 1287, [Erratum: 2007, JPCRD, 36, 1737]
Wiese, W. L., Fuhr, J. R., & Deters, T. M. 1996, Atomic Transition Probabilities of Carbon, Nitrogen, and Oxygen: A Critical Data Compilation

(J. Phys. Chem. Ref. Data Monograph No. 7) (New York: AIP), https://srd.nist.gov/JPCRD/jpcrdM7.pdf

Wolfire, M. G., McKee, C. F., Hollenbach, D., & Tielens, A. G. G. M. 2003, ApJ, 587, 278

Yan, Y., Taylor, K. T., & Seaton, M. J. 1987, JPhB, 20, 6399

Yerokhin, V. A., Surzhykov, A., & Müller, A. 2017, PhRvA, 96, 042505 [Erratum: 2017, PhRvA, 96, 069901]

Young, P. R., Feldman, U., & Lobel, A. 2011, ApJS, 196, 23

Zhang, Y., Liu, X.-W., Luo, S.-G., Péquignot, D., & Barlow, M. J. 2005, A&A, 442, 249